\documentstyle[12pt]{article}
\makeatletter \oddsidemargin  -.1in \evensidemargin -.1in
\newcommand{\singlespacing}{\let\CS=\@currsize\renewcommand{\baselinestretch}{1}\tiny\CS}
\newcommand{\oneandahalfspacing}{\let\CS=\@currsize\renewcommand{\baselinestretch}{1.25}\tiny\CS}
\newcommand{\doublespacing}{\let\CS=\@currsize\renewcommand{\baselinestretch}{1.35}\tiny\CS}

\newtheorem{rule-def}[theorem]{Rule}

\RequirePackage[dvips]{graphicx} \textheight 21.5cm

\begin{document}

%\singlespacing
%\doublespacing
\title{\bf Effect of Induced Magnetic Field on Peristaltic Flow of a Micropolar Fluid in an Asymmetric Channel}
\author{\small G. C. Shit$^a$\thanks{Email address: gcs@math.jdvu.ac.in
(G. ~C. ~Shit)}, ~~M. ~Roy$^a$ ~~and ~~E. ~Y. ~K. ~Ng$^b$
\thanks{Corresponding author. Email address: mykng@ntu.edu.sg
(E. Y. K. Ng)} \\
\it $^a$Department of Mathematics\\
\it Jadavpur University, Kolkata - 700032, India\\ \\
\it $^b$School of Mechanical and Aerospace Engineering,\\
\it College of Engineering,\\
\it Nanyang Technological University, \\
\it 50 Nanyang Avenue, Singapore - 639798 }
\date{}
\maketitle \noindent \doublespacing
\begin{abstract} Of concern in this paper is an investigation of
peristaltic transport of a physiological fluid in an asymmetric
channel under long wave length and low-Reynolds number
assumptions. The flow is assumed to be incompressible, viscous
electrically conducting micropolar fluid and the effect of induced
magnetic field is taken into account. Exact analytical solutions
obtained for the axial velocity, microrotation component, stream
line pattern, magnetic force function, axial induced magnetic
field as well as the current distribution across the channel. The
flow phenomena for the pumping characteristics, trapping and
reflux are also investigated. The results presented reveal that
the velocity decreases with the increase of magnetic field as well
as the coupling parameter. Moreover the trapping fluid can be
eliminated by the application of an external magnetic field. Thus
the study bears the promise of important applications in physiological
systems. \\

\end{abstract}

\section {Introduction:}
Peristaltic transport of a physiological fluid is induced by a
propagation of progressive waves of contraction or expansion along
the length of a distensible tube containing fluid. Peristaltic
mechanism encounters in urine transport from kidney to the
bladder, the movement of chyme in the gastrointestinal tract,
fluids in the lymphatic vessels, bile from the gallbladder into
the duodenum, the movement of spermatozoa in the ductus efferent
of the male reproductive tract, the movement of the ovum in the
fallopian tube and the circulation of blood in small blood
vessels. This mechanism also finds many applications in roller and
finger pumps, some bio-mechanical instruments, $e.g.$ heart-lung
machine, blood pump machine and dialysis machine. Thus,
peristaltic transport has been the recent studies of many
researchers/scientists owing to the above mentioned applications
in bio-mechanical engineering and bio-medical technology.\\

Latham \cite{R1} first attempted the study of the mechanism of
peristaltic transport and later on several theoretical and
experimental studies [2-10] have been made by others to understand
the clear idea of peristaltic action in different situations.
Srivastava and Srivastava \cite{R11} presented a mathematical
model by considering the effects of suspension of small spherical
rigid particles in an incompressible Newtonian viscous fluid in a
channel with flexible walls. Mekheimer et al. \cite{R12} analyzed
the mechanism of peristaltic pumping of a particle fluid
suspension in a planar channel. The peristaltic transport of a
Power-law fluid in the male reproductive tract was investigated by
Srivastava and Srivastava \cite{R13}. Usha and Rao \cite{R14}
carried out the study of peristaltic pumping of two-layered
Power-law fluids in a cylindrical tube. Lee and Fung \cite{R15}
studied theoretically the peristaltic transport phenomena in blood
flow, while Carew and Pedley \cite{R16} put forwarded a
mathematical
analysis for the study of peristaltic pumping in the ureter.\\

In all the above mentioned studies the peristaltic flows are
conducted in symmetric or axisymmetric channel or tube. Taylor
\cite{R17} conducted a theoretical study on asymmetric wave
propagation with an aim to explain the mechanical interaction
between neighboring spermatozoa. Mishra and Rao \cite{R18,R19}
made an investigation on peristaltic flow of a Newtonian fluid in
an asymmetric channel. Reddy et al. \cite{R20} considered the flow
of a Power-law fluid in an asymmetric channel caused by the
movement of peristaltic wave with the same speed but with
different amplitudes and phases on the flexible channel. Eytan and
Elad \cite{R21} have developed a mathematical model of
wall-induced peristaltic fluid flow in a channel with wave trains
having phase difference on the upper and lower walls. Most of
these investigations on peristaltic flow dealt with a Newtonian or
non-Newtonian fluid. However, it is observed that the complex
rheology of biological fluids behave as a suspensions of
deformable or rigid particle in a Newtonian fluid. For example,
blood is a suspension of various cells in plasma and cervical
mucus is a suspension of macromolecules in a water-like liquid.\\

Eringen \cite{R22} initiated the concept of micropolar fluids to
characterize the suspensions of neutrally buoyant rigid particles
in a viscous fluid. The micropolar fluids exhibit microrotational
and microintertial effects and support body couple and couple
stresses. It may be noted that micropolar fluids take care of the
microrotation of fluid particles by means of an independent
kinematic vector called microrotation vector. In view of this,
several investigators [23-28] carried out the study of the
peristaltic motion of micropolar fluid under different situations.
Ali and Hayat \cite{R29} investigated the study of peristaltic
motion of an incompressible micropolar fluid in an asymmetric
channel. Hayat and Ali \cite{R30} studied the effect of an
endoscope on peristaltic flow of a micropolar fluid through a
coaxial uniform tube, wherein they used long wave length approximation.\\

Recently, the study of magneto-hydrodynamic (MHD) flow of
electrically conducting fluids on peristaltic motion have become a
subject of growing interest for researchers and clinicians. This
is due to the fact that such studies are useful particularly for
pumping of blood and magnetic resonance imaging (MRI). Theoretical
work of Agarwal and Anwaruddin \cite{R31} explored the effect of
magnetic field on the flow of blood in atherosclerotic vessels of
blood pump during cardiac operations. Li et al. \cite{R32}
observed that an impulsive magnetic field can be used for a
therapeutic treatment of patients who have stone fragments in
their urinary tract. Makheimer and Al-Arabi \cite{R33} studied
non-linear peristaltic transport of physiological fluids through a
medium under the influence of a magnetic field. Misra et
al.\cite{R34} dealt with a theoretical investigation of the
peristaltic transport of a physiological fluid in a porous
asymmetric channel under the action of a magnetic field. Elshahed
and Harun \cite{R35} made an observation on the peristaltic
transport of Johnson -Segalman fluid by means of an infinite train
of sinusoidal waves in a flexible channel under the effect of a
magnetic field. However, all of these studies have neglected the
effect of induced magnetic field.\\

Vishnyakov and Pavlov \cite{R36} who first considered the effect
of induced magnetic field on peristaltic flow of an electrically
conducting Newtonian fluid. Lately, Makheimer \cite{R37, R38}
carried out the study of peristaltic transport of an
incompressible conducting micro-polar fluid in a symmetric channel
and another for a couple stress fluid by considering the effect of
induced magnetic field. In view of all the aforesaid observations,
the present analytical study has been designed in such a manner
that it can explore variety of information to study the effect of
induced magnetic field on peristaltic flow of a micro-polar fluid
in an asymmetric channel. The peristaltic wave train on the
channel wall has been considered to have different amplitudes and
phase difference. Thus, the results presented here will find an
important clinical applications such as in magnetic resonance
imaging (MRI) as well as in the gastrointestinal tract and small
blood vessels.
\section{Mathematical Formulation:}
Let us consider the unsteady flow of a viscous, incompressible and
electrically conducting micro-polar fluid through an asymmetric
channel of uniform thickness under the action of a magnetic field.
Let $Y'=h'_1$ and $Y'=h'_2$ be respectively the upper and lower
walls of the channel. The medium is considered to be induced by a
sinusoidal wave train propagating with a wave speed $c$ along the
channel wall (cf. Fig. 1), such that
\begin{eqnarray}
         h'_1(X',t')=d_1+a_1\cos\left[\frac{2\pi}{\lambda}(X'-ct')\right]
\end{eqnarray}
\begin{eqnarray}
         h'_2(X',t')=-d_2-a_2\cos \left[\frac{2\pi}{\lambda}(X'-ct')+\phi \right]
 \end{eqnarray}
     where $ a_1$ and $a_2 $ are the amplitudes of waves, $\lambda$ is the wave length,
 $\phi~(0\leq\phi\leq\pi)$ the phase difference, $X'$ and $Y'$ are the
 rectangular co-ordinates with $X'$ measured the axis of the channel and
 $Y'$ the traverse axis perpendicular to $X'$.\\
         The system is stressed by an external transverse uniform
constant magnetic field of strength $H'_0$, which will give rise
to an induced magnetic field $H'(h'_{X'},h'_{Y'},0)$ and therefore
the total magnetic field will be ${H'}^{+}(h'_{X'},H'_{0}
+h'_{Y'},0)$.\\
The equations of motion for unsteady flow of an incompressible
magneto-micro-polar fluid by neglecting the body couples are,
\begin{eqnarray}
   \vec{\nabla}\cdot \vec{ v'}=0
\end{eqnarray}
\begin{eqnarray}
        \rho\left (\frac{\partial \vec{v'}}{\partial t'}
   +(\vec{v'}\cdot\vec{\nabla})\vec{v'}\right )=-\nabla \left (p'+
   \frac{1}{2}\mu_e(H'^{+})^2 \right )+(\mu+k)\nabla^{2}\vec{v'}+
   k\vec{\nabla}\times\vec{\omega'}-
   \mu_e(\vec{H'}^{+}\cdot\vec{\nabla})\vec{H'}^{+}
\end{eqnarray}
\begin{eqnarray}
         \rho J\left (\frac{\partial \vec{w'}}{\partial t'}
   +(\vec{v'}\cdot\vec{\nabla})\vec{\omega'}\right )=-2k\vec{\omega'}
   +k\vec{\nabla}\times \vec{v'}-
   \gamma(\vec{\nabla}\times\vec{\nabla}\times\vec{\omega'})+
   (\alpha+\beta+\gamma)\vec{\nabla}(\vec{\nabla}\cdot\vec{\omega'})
\end{eqnarray}
where, $\nabla^{2}=\frac{\partial^{2}}{\partial X'^{2}}+
      \frac{\partial^{2}}{\partial Y'^{2}}$

and  $\vec{v'}$ is the velocity vector, $\vec{\omega'}$ is the
 microrotation vector, $p'$ is the fluid pressure, $\rho$ the fluid density, $J$ the micro-gyration
 parameter,
 $\sigma$ the electrical conductivity, $J'$ is the electrical
 current density, $\mu_{e}$ is the magnetic permeability,
 $\vec{E'}$ is an induced electrical field. Also the material constants
(or viscosity coefficients of the micro-polar fluid ) $\mu$, $k$,
$\alpha$, $\beta$, and $\gamma$ are satisfies the following
inequalities

           $(2\mu+k)\geq 0$, $k\geq 0$,$(3\alpha+\beta+\gamma)\geq 0$, $\gamma\geq |\beta |$

 \noindent Along with the Maxwell's equations,
\begin{eqnarray}
         ~~ \vec{\nabla}\times \vec{H'}=\vec{J'},~~~~~~~~~~
       \nabla\times \vec{E'}=-\mu_{e}\left (\frac{\partial
       \vec{H'}}{\partial t'}\right )
\end{eqnarray}
   and with Ohm's law:
\begin{eqnarray}
       \vec{J'}=\sigma \left (\vec{E'}+\mu_{e}(\vec{V}\times \vec{H'}^{+})\right )
\end{eqnarray}
in addition, it should be noted that
\begin{eqnarray}
\nabla\cdot \vec{H'}=0, ~~\vec{\nabla}\cdot \vec{E'}=0,
\end{eqnarray}

\noindent Now, combining equations (6) - (8) we get the induction
equation,
\begin{eqnarray}
        \frac{\partial \vec{H'}^{+}}{\partial t'}=
        \vec{\nabla}\times(\vec{v'}\times \vec{H'}^{+})
        +\frac{1}{\mu_{e}\sigma} \nabla^{2}\vec{H'}^{+}
\end{eqnarray}
   Let us consider a wave frame $(x',y')$ moving with the
   velocity $c$ away from fixed frame $(X',Y')$ have
   the transformations
\begin{eqnarray}
                 x'=X'-ct',~y'=Y',~
                 u'=U'-c,~ v'=V',
\end{eqnarray}
  \noindent where $(U',V')$ and $(u',v')$ are the component of velocity in
 the fixed frame and wave frame of reference respectively.\\
\noindent Now, we introduce the dimensionless variables as
follows:
\begin{eqnarray}
                            x=\frac{x'}{\lambda},~
                            y=\frac{y'}{a},~
                            u=\frac{u'}{c},~ v=\frac{\lambda
                            v'}{ac},~
                            h(x)=\frac{h'(x')}{a},~
                            p=\frac{a^{2}p'(x')}{\lambda \mu c}~
                            t=\frac{ct'}{\lambda},\nonumber\\
                            J=\frac{J'}{a^{2}},~
                            \psi=\frac{\psi'}{ca},~
                            \phi=\frac{\phi'}{H_{0}a}
\end{eqnarray}
\noindent where $\psi$ and $\phi$ represent the stream function
and magnetic
force function respectively.\\
 Use of (11) in the equations (4)-(9) yields the MHD flow for the micropolar fluid
 in terms of the stream function $\psi(x,y)$ and magnetic force function $\phi(x,y)$ are
\begin{eqnarray}
                 R_{e}\delta \left [\left (\psi_{y}\frac{\partial}{\partial x}-
                 \psi_{y}\frac{\partial}{\partial y}\right )\psi_{y}\right ]=-\frac{\partial p_m}{\partial
                 x}+\frac{1}{1-N}\nabla^{2}\psi_y+\frac{N}{1-N}\frac{\partial\omega}{\partial
                 y}+R_{e}S^{2}\phi_{yy}\nonumber \\
                 +R_{e}S^{2}\delta \left [\left (\phi_y\frac{\partial}{\partial x}-
                 \phi_y\frac{\partial}{\partial y}\right
                 )\phi_y\right ]
\end{eqnarray}
\begin{eqnarray}
            R_{e}\delta^{3}\left [\left (\psi_{x}\frac{\partial}{\partial y}-
                 \psi_{y}\frac{\partial}{\partial x}\right )\psi_{x}\right ]=-\frac{\partial p_m}{\partial
                 y}+\frac{\delta^{2}}{1-N}\nabla^{2}\psi_x+\frac{\delta^{2}N}{1-N}\frac{\partial\omega}{\partial
                 x}+R_{e}S^{2}\phi_{xy}\nonumber \\
                 -R_{e}S^{2}\delta^{3}\left [\left (\phi_y\frac{\partial}{\partial x}-
                 \phi_y\frac{\partial}{\partial y}\right )\phi_x\right
                 ]
\end{eqnarray}
\begin{eqnarray}
           R_{e}\delta J\frac{1-N}{N}\left [\left (\psi_{y}\frac{\partial}{\partial x}-
                 \psi_{y}\frac{\partial}{\partial
                 y}\right )\omega \right ]=-2\omega-\nabla^{2}\psi+\frac{2-N}{N}\nabla^{2}\omega
\end{eqnarray}
\begin{eqnarray}
           \psi_{y}-\delta
           (\psi_{x}\phi_{y}-\psi_{y}\phi_{x})+\frac{1}{R_{m}}\nabla^{2}\phi=E
\end{eqnarray}
                 with
\begin{eqnarray}
                 u=\frac{\partial\psi}{\partial y},~ v=-\delta\frac{\partial\psi}{\partial
                 x},~h_{x}=\frac{\partial \phi}{\partial y},~
                 h_{y}=-\delta\frac{\partial\phi}{\partial x},~
                 \nabla^{2}=\delta^{2}\frac{\partial^{2}}{\partial
                 x^{2}}+\frac{\partial^{2}}{\partial y^{2}}
\end{eqnarray}
  We define the following dimensionless parameters that appear in (12) - (16) as \\
     Reynolds number $R_{e}=\frac{ca\rho}{\mu}$\\
     Wave number $\delta=\frac{a}{\lambda}$\\
     Strommer's number (magnetic force number)
            $S=\frac{H_{0}}{c}\sqrt{(\frac{\mu_{e}}{\rho})}$\\
     The magnetic Reynolds number $R_{m}=\sigma\mu_{e}ac$\\
     The coupling number $N=\frac{k}{k+\mu}$, $(0\leq N \leq
    1)$,\\ and $m^{2}=\frac{a^{2}k(2\mu+k)}{(\gamma(\mu+k))}$ is
    the micropolar or microrotation  parameter. \\
     The total pressure in the fluid, which is equal to the sum
    of the ordinary and magnetic pressure given by,
    $P_{m}=P+\frac{1}{2}R_{e}\delta(\frac{\delta (H^{+})^{2}}{\rho
    c^{2}}$, and $E(=-\frac{E}{\mu_ecH_{0}})$ is defined as the electrical field
    strength.\\
         By eliminating the pressure term between (12) and (13) we obtain,

\begin{eqnarray}
               R_{e}\delta \left [\left (\psi_{y}\frac{\partial}{\partial x}-
                 \psi_{y}\frac{\partial}{\partial
                 y}\right )\nabla^{2}\psi \right ]=\frac{1}{1-N}\nabla^{4}\psi+\frac{N}{1-N}\nabla^{2}\omega \nonumber \\
                 +R_{e}S^{2}\nabla^{2}\phi_{y}+R_{e}S^{2}\delta \left [\left (\phi{y}\frac{\partial}{\partial x}-
                 \phi{x}\frac{\partial}{\partial y}\right )\nabla^{2}\phi \right]
\end{eqnarray}
 with an aim to test the validity of our present analysis, we made an
 assumption that when $k \longrightarrow 0$, the equations (4) and (5)
 reduces to the classical MHD Navier-stokes equations.

\noindent The volumetric flow rate in the fixed frame is given by
\begin{eqnarray}
          Q=\int^{h'_{1}}_{h'_{2}}U'(X',Y',t')dy'
\end{eqnarray}
                  where $h'_{1}$ and $h'_{2}$ are functions of $X'$ and $t'$.\\
The rate of volume flow in the wave frame is found to be given by,
\begin{eqnarray}
             q=\int^{h'_{1}}_{h'_{2}}u'(x',y')dy'
\end{eqnarray}
 where $h'_{1}$, and $h'_{2}$  are functions of $X'$ alone.\\
 The dimensionless form of the peristaltic channel walls are
\begin{eqnarray}
                 h_{1}(x)=1+a\cos(2\pi x),~~~~~~
                 h_{2}(x)=-d-b\cos(2\pi x+\phi)
\end{eqnarray}
where,
                $ a=\frac{a_{1}}{d_{1}}$,~~
                 $b=\frac{a_{2}}{d_{1}}$,~~$d=\frac{d_{2}}{d_{1}}$.
                 \\
 Putting (10) into the equations (18) and (19), the relation between {\it Q} and {\it q} can be obtained as
 \begin{eqnarray}
           Q=q+c(h'_{1}-h'_{2})
\end{eqnarray}
The time mean flow over a period T at a fixed position $X'$ is
defined as
\begin{eqnarray}
                  Q'=\frac{1}{T}\int^{T}_{0}Q dt
\end{eqnarray}
Using (21) in (22) the flow rate $Q'$ has the form,
\begin{eqnarray}
           Q'=\frac{1}{T}\int^{T}_{0}Q dt +c(h'_{1}-h'_{2})
                  =q+cd_{1}+cd_{2}
\end{eqnarray}
 The non-dimensional form of (23) is given by
\begin{eqnarray}
                 \theta=F+1+d
\end{eqnarray}
 where $\theta=\frac{Q'}{cd_{1}}$, $F=\frac{q}{cd_{1}}$
 and $d=\frac{d_{2}}{d_{1}}.$\\
  In which,
\begin{eqnarray}
                F=\int^{h_{1}}_{h_{2}}\frac{\partial\psi}{\partial
                y}dy=\psi(h_{1})-\psi(h_{2})
\end{eqnarray}

The boundary conditions for the dimensionless stream function
$\psi(x,y)$ and magnetic force function $\phi(x,y)$ in the wave
frame is considered as
\begin{eqnarray}
                       \frac{\partial\psi}{\partial y}=-1,   ~ on~  y=h_{1}~and ~y=h_{2}\nonumber\\
                       \omega=0,                 ~on ~y=h_{1} ~and ~y=h_{2}\nonumber\\
                       \psi=\frac{F}{2},  ~ on ~ y=h_{1}\nonumber\\
                           =-\frac{F}{2},  ~on  ~y=h_{2}\nonumber\\
                   and~ \phi=\frac{\partial\phi}{\partial y}=0 ~on ~ y=h_{1}~ and ~y=h_{2}\nonumber\\
\end{eqnarray}
 Under the assumption of long wave length and low Reynolds
 number, the dimensionless equations (17), (14) and (15) becomes
 \begin{eqnarray}
               \frac{\partial^{4}\psi}{\partial y^{4}}+N\frac{\partial^{2}\omega}{\partial
               y^{2}}+R_{e}S^{2}(1-N)\frac{\partial^{3}\phi}{\partial
               y^{3}}=0
\end{eqnarray}
\begin{eqnarray}
              \frac{(2-N)}{m^{2}}\frac{\partial^{2}\omega}{\partial
              y^{2}}=2\omega+\frac{\partial^{2}\psi}{\partial y^{2}}
\end{eqnarray}
\begin{eqnarray}
                \frac{\partial^{2}\phi}{\partial y^{2}}=R_{m}(E-\frac{\partial\psi}{\partial
                y})
\end{eqnarray}
By operating $\frac{\partial^2}{\partial y^2}$ and
$\frac{\partial}{\partial y}$ on both sides of (28) and (29)
respectively and substituting the expressions for
$\frac{\partial^{4}\psi}{\partial y^{4}}$ and
  $\frac{\partial^{3}\phi}{\partial y^{3}}$ in equation (27) we obtain as
\begin{eqnarray}
                  \frac{\partial^{4}\omega}{\partial y^{4}}-\{m^{2}+
                  (1-N)H^{2}\}\frac{\partial^{2}\omega}{\partial^{2}y}+
                  \frac{2m^{2}H^{2}(1-N)}{(2-N)}\omega=0
\end{eqnarray}
Again, substituting the value of
$\frac{\partial^{2}\omega}{\partial y^{2}}$
 from (28) and $\frac{\partial^{3}\phi}{\partial y^{3}}$ from
 (29) after differentiating with respect to y, in (27) the expression for stream function may be written as
\begin{eqnarray}
        \psi=\frac{1}{H^{2}(1-N)}\left[{\frac{(2-N)}{m^{2}}\{\frac{\partial^{2}\omega}{\partial
        y^{2}}-m^{2}\omega\}+C_{1}y+C_{2}}\right]
\end{eqnarray}
                  where $H^{2}=R_{e}S^{2}R_{m}$~ $\Rightarrow H=(\mu_{e}H_{0})\sigma\sqrt{\frac{\sigma}{\mu}}$
   is the Hartmann number $(>\sqrt{2})$ and $C_{1}$ and $C_{2}$ are two integrating constants.\\
Then the general solution of (30) takes the form,
\begin{eqnarray}
               \omega=A\cosh(\theta_{1}y)+B\sinh(\theta_{1}y)+C\cosh(\theta_{2}y)+D\sinh(\theta_{2}y),
\end{eqnarray}
where\\
 ~~~~~~~~~ $\theta_{1}=\frac{1}{\sqrt{2}}\sqrt{\{(1-N)H^{2}+m^{2}\}+\sqrt{\{(1-N)H^{2}+m^{2}\}^{2}-\frac{8m^{2}(1-N)H^{2}}{(2-N)}}}$\\
 and
 ~~~~~~~~~ $\theta_{2}=\frac{1}{\sqrt{2}}\sqrt{\{(1-N)H^{2}+m^{2}\}-\sqrt{\{(1-N)H^{2}+m^{2}\}^{2}-\frac{8m^{2}(1-N)H^{2}}{(2-N)}}}$\\
together with four integrating constants A, B, C and D. \\
  Using the expression for $\omega$ from (32) and (31), the
  general solution for $\psi$ is given by
\begin{eqnarray}
\psi&=&\frac{1}{H^{2}(1-N)}
\Big [\frac{(2-N)}{m^{2}}\{(\theta^{2}_{1}-m^{2})(A\cosh(\theta_{1}y)+B\sinh(\theta_{1}y))\nonumber \\
&+&(\theta^{2}_{2}-m^{2})(C\cosh(\theta_{2}y)+D\sinh(\theta_{2}y))\}+C_{1}y+C_{2}
\Big ]
\end{eqnarray}
   Applying the boundary conditions given in the equation (26) into the equations (32) and (33),
    the values of all the constants have
    been determined and are presented in the appendix. \\
Thus, the expressions for the axial velocity $u$ and stream
function $\psi$ are obtained as
\begin{eqnarray}
                u=\frac{\partial \psi}{\partial
                y}=\frac{1}{Z_{0}}\left[Z_{1}\{A\sinh(\theta_{1}y)+B\cosh(\theta_{1}y)\}
                +Z_{2}\{C\sinh(\theta_{2}y)+D\cosh(\theta_{2}y)\}+C_{1}\right]
\end{eqnarray}
and
\begin{eqnarray}
      \psi(x,y)=\frac{1}{Z_{0}}\left[\frac{Z_{1}}{\theta_{1}}\{A\cosh(\theta_{1}y)+B\sinh(\theta_{1}y)\}
      +\frac{Z_{2}}{\theta_{2}}\{C\cosh(\theta_{2}y)+D\sinh(\theta_{2}y)\}+C_{1}y+C_{2}\right]
\end{eqnarray}
Substituting the value of $\psi$ obtained from (35) in (29) and
integrating we get the magnetic force function $\phi$ as,
\begin{eqnarray}
 \phi(x,y)=R_{m}E\frac{y^2}{2}-\frac{R_{m}}{Z_{0}}\left[\frac{Z_{1}}{\theta^{2}_{1}}\{A\sinh(\theta_{1}y)+B\cosh(\theta_{1}y)\}\right.\nonumber\\\left.
           +\frac{Z_{2}}{\theta^{2}_{2}}\{C\sinh(\theta_{2}y)+D\cosh(\theta_{2}y)\}+C_{1}\frac{y^{2}}{2}+C_{2}\right]+C_{3}y+C_{4}
\end{eqnarray}
where the constants $C_{3}$ and $C_{4}$ are obtained so far using
the boundary conditions (26) are also included in the appendix.\\
The axial-induced magnetic field is given by,
\begin{eqnarray}
      h_{x}=\frac{\partial\phi}{\partial y}=R_{m}E y-R_{m}\psi+C_{3}
\end{eqnarray}
and the current density distribution across the channel has the
form,
\begin{eqnarray}
J_{z}=R_{m}E-R_{m}u
\end{eqnarray}
The electric field strength E is determined by integrating (29)
and taking the boundary conditions on $\phi$~ and ~$\psi$~across
the wall surface we get the dimensionless form as
\begin{eqnarray}
 E=\frac{F}{h_{1}(x)-h_{2}(x)}
\end{eqnarray}
In order to obtain the pumping characteristics by means of
pressure rise per wavelength the axial pressure gradient can be
determined from the equation
\begin{eqnarray}
 \frac{\partial p}{\partial x}=\frac{1}{(1-N)}\frac{\partial^{3}\psi}{\partial
 y^{3}}+\frac{N}{(1-N)}\frac{\partial \omega}{\partial y}+H^{2}(E-\frac{\partial \psi}{\partial y})
\end{eqnarray}
which imply,
\begin{eqnarray}
   \frac{\partial p}{\partial
   x}=\frac{1}{z_{0}(1-N)}\{Z_{1}\theta^{2}_{1}(A\sinh(\theta_{1}y)+B\cosh(\theta_{1}y))
       +Z_{2}\theta^{2}_{2}(C\sinh(\theta_{2}y)+D\cosh(\theta_{2}y))\}\nonumber\\
       +\frac{N}{(1-N)}\{A\theta_{1}(sinh(\theta_{1}y)+B\theta_{1}\cosh(\theta_{1}y)+C\theta_{2}\sinh(\theta_{2}y)+D\theta_{2}\cosh(\theta_{2}y)\}\nonumber\\
       +H^{2}(E-u)
\end{eqnarray}
The non-dimensional expression of pressure rise per wave length
$\triangle P$~ is given by,
\begin{eqnarray}
      \triangle p=\int^{1}_{0}\frac{\partial p}{\partial x}dx
\end{eqnarray}
It is interesting to note that the stress tensor in micropolar
fluid is not symmetric. Therefore, the dimensionless form of the
shear stress involved in the present problem under consideration
are given by
\begin{eqnarray}
    \tau_{xy}=\frac{\partial u}{\partial y}-\frac{N}{(1-N)}\omega
\end{eqnarray}
\begin{eqnarray}
\tau_{yx}=\frac{1}{(1-N)}\frac{\partial u}{\partial
y}+\frac{N}{(1-N)}\omega
\end{eqnarray}
The numerical computations for the shear stresses $\tau_{xy}$ and
 $\tau_{yx}$ are obtained at both the upper and lower walls of the
channel and whose graphical representation is presented in the
next section.

\section{Computational Results and Discussion:}
In order to examine the flow characteristics of peristaltic
transport of a magneto-micro-polar fluid by taking into account
the effect of induced magnetic field, the analytical expressions
for various quantities of interest have been analyzed in the
previous section. For the purpose of numerical computation it is
necessary to assign some valid numerical values to the physical
parameters involved in the problem. In this connection, the
following experimental data of physiological fluids or foregoing
theoretical analysis have been used:
 $a=0.3, ~0.5; ~b=0.3, ~0.5; ~d=1.0; ~H=2, ~4, ~6, ~8; ~N=0.2, ~0.4, ~0.6, ~0.8;
 ~m=0.1, ~5, ~10, ~100$; $\theta=2.4$,~ $\phi=0.0$,  $\frac{\pi}{2}$,
 $\frac{5\pi}{6}$,~$\pi$~~and a low magnetic Reynolds number
 {\it$ R_m=1.0$} (cf. Fig. 1).\\
Figs. 2-5 illustrate the distribution of axial velocity $u$ along
the height of the channel for different values of the
dimensionless parameters. The variations of Hartmann number $ H$
on axial velocity $u$ are shown in Figs. 2 and 3 with positive and
negative flow rate respectively. The mean velocity change is occur
in the direction opposite to the direction of net flow, refers to
reflux when $\theta=2.4$. This phenomena is of considerable
importance in some physiological fluids. But it has been observed
that there is no reflux for~~ $\theta<-1.5$~~ The reversal flow
reduces as the Hartmann number $H$ increases. The effect of
coupling number $N$ on axial velocity is shown in Fig. 4. It is
noticed that the flow reversal near the central line of the
channel increases with the increase of coupling number, while the
reversal trend is occur in the vicinity of the channel walls. But
no significant change is found for the variation of micro-polar
parameter $m$. When the phase difference $\phi$ increases, the
axial velocity near the lower wall of the channel is increases,
whereas the velocity is constant near the upper wall of the
channel (cf. Fig. 5). This is due to the imposition of phase
difference in the walls of the channel by means of an asymmetry,
which causes the elimination of flow reversal.\\

\noindent Figs.6-8 give the variation of pressure rise per wave
length with dimensionless volumetric flow rate $\theta$ for
different values of the Hartmann number $H$, microrotation
parameter m, as well as the phase difference $\phi$. There are
three region of peristaltic pumping viz, pumping region for
$\triangle P>0$ with $\theta>0$ (positive pumping) and $\theta<0$
(negative pumping); free pumping region for $\triangle P=0$ and
co-pumping region for $\triangle P<0$. It may observe from Fig. 6
that in the pumping region with positive pumping, the pressure
rise $\triangle P$ as well as the rate of flux increases and it
decreases in the co-pumping region with the increase of Hartmann
number $H$. Fig. 7 shows that in the entire co-pumping region the
pressure rise per wave length and the volume flux $\theta$
increases with increasing phase difference $\phi$. Moreover, same
behaviour is observed in Fig. 8 in the case of microrotation
parameter m butthe change is insignificant.\\

\noindent The variations of the axial pressure gradient
$\frac{\partial p}{\partial x}$ over one wave length is presented
through the Figs. 9-11. It reveals from Figs. 9 and 10 that in the
wider part of the channel when $x\epsilon[0.0,0.2]$ and
$x\epsilon[0.8,1.0]$ the pressure gradient is relatively small,
where the flow can easily pass without giving any large pressure
gradient. On the other hand, a large pressure gradient is required
to maintain the same flux to pass it in a narrow part of the
channel for $x\epsilon[0.2,0.8]$. It is observed from Fig. 11 that
the magnitude of the pressure gradient is reduced due to the
increase of phase difference $\phi$. Thus the phase shift helps to
pass same flux with imposing less pressure gradient. Also it has
been noticed from Figs. 9 and 10 that the magnitude of the
pressure gradient increases with the increase of both Hartmann
number $H$ and coupling parameter N.\\

\noindent The most important characteristics in the present
problem are the effect of induced magnetic field subject to the
action of an externally applied magnetic field $H_{0}$. The effect
of axial induced magnetic field $h_{x}$ along the height of the
channel is presented through Figs. 12 and 13 for a particular
value of $x=1$. It is seen from Figs. 12 that the magnitude of the
induced magnetic field decreases on the upper half of the channel
whereas it increases on the lower half of the channel with the
increase of $H$. It is interesting to note from Fig. 13 that the
induced magnetic field increases near the lower half of the
channel with increasing phase difference $\phi$. Thus an asymmetric
channel causes an increase in axial induced magnetic field.\\

\noindent The distribution of current density $J_{z}$ are shown in
Figs. 14 and 15. The magnitude of the current density $J_{z}$
decreases near the central line of the channel with the increasing
Hartmann number $H$, while the trend is reversed near the
boundaries. One may observe from Fig. 15 that the current density
decreases with the increase of the phase difference $\phi$ near
the lower wall of the channel, however in the upper wall $J_{z}$
remains constant.\\

It is well known that the stress tensor is not symmetric in the
micro-polar fluid whose analytical expressions presented in
equations (43) and (44). The shear stress $\tau_{xy}$ and
$\tau_{yx}$ on both the upper and lower wall of the channel for
different values of the various parameters illustrated in Figs.
16-19. The variation of the Hartmann number $H$ on the shear
stress $\tau_{xy}$ is shown in Fig. 16 and which reveals that the
shear stress increases at the upper wall and decreases at the
lower wall of the channel with an increase in $H$. For the sake of
comparison we have also examined for the shear stress $\tau_{yx}$
with different values of $H$. We observed that similar is the
variation for $\tau_{yx}$ as in the case of $\tau_{xy}$ and that
they are the same but differ only by magnitude. However, there is
much difference in variation of coupling parameter $N$ on the
shear stress $\tau_{xy}$ and $\tau_{yx}$ presented in Figs. 17 and
18. These two figures show that the shear stress $\tau_{xy}$
decreases on the upper wall and increases on the lower wall of the
channel as $N$ increases. On the other hand $\tau_{yx}$ increases
at the upper wall and decreases on the lower wall when {\it N}
increases. It has been observed from Fig. 19 that the pick value
of the shear stress $\tau_{xy}$ decreases with the increase of
phase difference $\phi$ in the case of upper wall whereas in the
case of lower wall the change is reversed. It has also been
examined for the variation of microrotation parameter $m$ and
phase difference $\phi$ on the shear stress $\tau_{yx}$. It is to
be believed that similar is the observation for the variation of
different values of $m$ and $\phi$ on the shear stress $\tau_{yx}$
on both the walls as that of the variation of $\tau_{xy}$. \\

\noindent {\bf Trapping Phenomena:} Shapiro et al [39] held that
at high flow rates and large occlusions there is a region of
closed streamlines in the wave frame and thus some fluid is found
trapped within a wave propagation. The trapped fluid mass was
found to move with a same speed equal to the wave motion. Thus,
the phenomenon of trapping may be looked upon as the formation of
an internally circulating bolus of fluid. Owing to the trapping
phenomenon, there will exist stagnation points, where both the
velocity components of the fluid vanish in the wave frame. Figs.
20-22 illustrate the effect of Hartmann number $H$ on the
streamlines for peristaltic transport of a micropolar fluid. These
figures show that under the action of low magnetic field, the
bolus is formed near the central region of the channel. It is
interesting to note that the bolus is decreases in size and
ultimately vanishes as the Hartmann number {\it H} increases.
Thus, the stagnation points will be located at the central region
of the channel and finally it can be eliminated by the application
of strong magnetic field. The influence of coupling parameter {\it
N} on the streamlines is demonstrated in Figs. 23-25. It has been
observed that as the coupling parameter $N$ increases the bolus
increases in size and formed more inner and inner circulations. A
very interesting fact shown in Figs. 26-28 that the bolus appears
in the central region for $\phi=0$ (cf. Fig. 20) moves towards the
left and decreases in size as $\phi$ increases. When $\phi=\pi$,
the bolus completely vanishes and the streamlines are aligned
parallel to one another.\\

\noindent Figs. 29-36 give the distribution of magnetic force
function $\phi(x,y)$ for different values of the Hartmann number
{\it H}, coupling parameter {\it N}, and phase difference $\phi$.
It has been shown that in the presence of low magnetic field, the
magnetic lines strongly appear in the central region of the
channel and are vanish rapidly as the Hartmann number {\it H}
increases. One may observe from Figs. 32-33 that the magnetic
force lines disappear from the boundaries of the channel when the
coupling parameter $N$ increases. But the magnetic lines remain
fixed at the central line of the channel. It is interesting to
note from Figs. 34-36 that for $\phi=0$ (cf. Fig. 29), the
magnetic lines in closed form appears once at the central region
and the closed form of magnetic lines broken up into two or more
closed form and moves backward to each other.\\

\section{Conclusions:}
In this paper, we have theoretically studied the effects of
induced magnetic field on peristaltic transport of physiological
fluid represented by micropolar fluid model in an asymmetric
channel. In this investigation, special emphasis has been paid to
study the flow features, the pumping characteristic, the trapping
phenomena and the axial induced magnetic force function. From the
presented analysis, one can make an important conclusion that it
is possible to increase pumping as often as necessary the pressure
gradient by applying an external magnetic field and that the bolus
formation can be eliminated by suitably adjusting the magnetic
field intensity. For an asymmetric channel i.e., due to the phase
differences at the channel walls, the flow can easily pass with
imposing large amount of pressure gradient. There is a linear
relationship between pressure rise per wavelength and volumetric
flow rate. In the entire co-pumping region, the pressure rise and
the volumetric flow rate both increases with the increase of phase
difference. The wall shear stress $\tau_{xy}$ and $\tau_{yx}$ both
increases with the increase of Hartmann number {\it H} at the
upper wall, whereas at the lower wall it decreases. The trapped
bolus can be eliminated by the application of an external magnetic
field. It may interesting to note that the trapped bolus can also
be eliminated due to the phase difference (channel asymmetry) at
the walls. When the coupling parameter increases, the trapped
bolus appears at the central region of the channel. The magnetic
force lines decrease from the central region and in the vicinity
of the channel walls with the increase of the Hartmann number {\it
H} and coupling parameter {\it N} respectively. Thus the results
presented through some light on  problems associated with fluid
movement in the gastrointestinal tract, intra-uterine fluid motion
induced by uterine contraction, as well as flow through small
blood vessels and intrapleural membranes.

{\bf Acknowledgements:} {\it The authors sincerely thank the
referees for their very useful suggestion based upon which the
original manuscript has been revised to the present form. The
authors G. C. Shit and M. Roy are thankful to CSIR, New Delhi for
the financial support during this investigation.}

\section{Appendix:}
The expressions that appear in section 2 are listed as follows:\\
$A=\frac{Z_{21}(Z_{20}-Z_{23})}{Z_{19}Z_{23}-Z_{20}Z_{22}}$,~~~~~~~~$B=\frac{Z_{21}(Z_{22}-Z_{19})}{Z_{19}Z_{23}-Z_{20}Z_{22}}$,~~~~~~$C=\frac{AZ_{14}+BZ_{15}}{Z_{16}}$,\\
$D=\frac{AZ_{17}+BZ_{18}}{Z_{16}}$,~~~~~~~$C_{1}=Z_{6}-AZ_{7}-BZ_{8}$~~~~~~$C_{2}=AZ_{11}+BZ_{12}-Z_{13}$\\
$C_{3}=-\frac{R_{m}E(h_{1}+h_{2})}{2}+\frac{R_{m}}{Z_{0}(h_{1}-h_{2})}[\frac{Z_{1}}{\theta^{2}_{1}}(AZ_{5}+BZ_{4})+\frac{Z_{2}}{\theta^{2}_{2}}(CZ_{24}+DZ_{25})]+\frac{R_{m}}{Z_{0}}[C_{1}\frac{(h^{1}+h_{2})}{2}+C_{2}]$,\\
$C_{4}=\frac{R_{m}Eh_{1}h_{2}}{2}-\frac{R_{m}}{Z_{0}(h_{1}-h_{2})}[\frac{Z_{1}}{\theta^{2}_{1}}(AZ_{10}+BZ_{9})+\frac{Z_{2}}{\theta^{2}_{2}}(CZ_{26}+DZ_{27})]+\frac{R_{m}C_{1}h^{1}h_{2}}{2Z_{0}}$,\\
        $\xi_{1}=\theta^{2}_{1}-m^{2}$,~~~$\xi_{2}=\theta^{2}_{2}-m^{2}$,\\
        $Z_{0}=H^2(1-N)$,~~~$Z_{1}=\frac{(2-N)\xi_{1}\theta_{1}}{m^2}$,~~~$Z_{1}=\frac{(2-N)\xi_{1}\theta_{1}}{m^2}$,~~~$Z_{3}=\frac{(2-N)(\xi_{1}-\xi_{2})}{m^{2}(h_{1}-h_{2})}$,\\
        $Z_{4}=\cosh(\theta_{1}h_{1})-\cosh(\theta_{1}h_{2})$,~~~$Z_{5}=\sinh(\theta_{1}h_{1})-\sinh(\theta_{1}h_{2})$,\\
        $Z_{6}=\frac{FZ_{0}}{h_{1}-h_{2}}$,~~~$Z_{7}=Z_{3}Z_{4}$,~~~$Z_{8}=Z_{3}Z_{5}$,\\
        $Z_{9}=h_{2}\cosh(\theta_{1}h_{1})-h_{1}\cosh(\theta_{1}h_{2})$,\\
        $Z_{10}=h_{2}\sinh(\theta_{1}h_{1})-h_{1}\sinh(\theta_{1}h_{2})$,\\
        $Z_{11}=Z_{3}Z_{9}$,~~~$Z_{12}=Z_{3}Z_{10}$,~~~$Z_{13}=\frac{Z_{6}(h_{1}+h_{2})}{2}$,\\
        $Z_{14}=\sinh(\theta_{2}h_{2})\cosh(\theta_{1}h_{1})-\cosh(\theta_{1}h_{2})\sinh(\theta_{2}h_{1})$,\\
        $Z_{15}=\sinh(\theta_{2}h_{2})\sinh(\theta_{1}h_{1})-\sinh(\theta_{1}h_{2})\sinh(\theta_{2}h_{1})$,\\
        $Z_{16}=\sinh[\theta_{2}(h_{1}-h_{2})]$,\\
        $Z_{17}=\cosh(\theta_{1}h_{2})\cosh(\theta_{2}h_{1})-\cosh(\theta_{1}h_{1})\cosh(\theta_{2}h_{2})$,\\
        $Z_{18}=\sinh(\theta_{1}h_{2})\cosh(\theta_{2}h_{1})-\cosh(\theta_{2}h_{2})\sinh(\theta_{1}h_{1})$,\\
        $Z_{19}=Z_{1}\sinh(\theta_{1}h_{1})+\frac{Z_{2}Z_{14}}{Z_{16}}\sinh(\theta_{2}h_{1})+\frac{Z_{2}Z_{17}}{Z_{16}}\cosh(\theta_{2}h_{1})-Z_{7}$\\
        $Z_{20}=Z_{1}\cosh(\theta_{1}h_{1})+\frac{Z_{2}Z_{15}}{Z_{16}}\sinh(\theta_{2}h_{1})+\frac{Z_{2}Z_{18}}{Z_{16}}\cosh(\theta_{2}h_{1})-Z_{8}$\\
        $Z_{21}=-Z_{0}-Z_{6}$,\\
        $Z_{22}=Z_{1}\sinh(\theta_{1}h_{2})+\frac{Z_{2}Z_{14}}{Z_{16}}\sinh(\theta_{2}h_{2})+\frac{Z_{2}Z_{17}}{Z_{16}}\cosh(\theta_{2}h_{2})-Z_{7}$\\
        $Z_{23}=Z_{1}\cosh(\theta_{1}h_{2})+\frac{Z_{2}Z_{15}}{Z_{16}}\sinh(\theta_{2}h_{2})+\frac{Z_{2}Z_{18}}{Z_{16}}\cosh(\theta_{2}h_{2})-Z_{8}$\\
$Z_{24}=\sinh(\theta_{2}h_{1})-\sinh(\theta_{2}h_{2})$,~~~~ $Z_{25}=\cosh(\theta_{2}h_{1})-\cosh(\theta_{2}h_{2})$,\\
$Z_{26}=h_{2}\sinh(\theta_{2}h_{1})-h_{1}\sinh(\theta_{2}h_{2})$,~~~~  $Z_{27}=h_{2}\cosh(\theta_{2}h_{1})-h_{1}\cosh(\theta_{2}h_{2})$,\\

\newpage
\textheight 22.0cm \pagebreak
\begin{minipage}{1.0\textwidth}
   \begin{center}
      \includegraphics[width=3.8in,height=2.5in ]{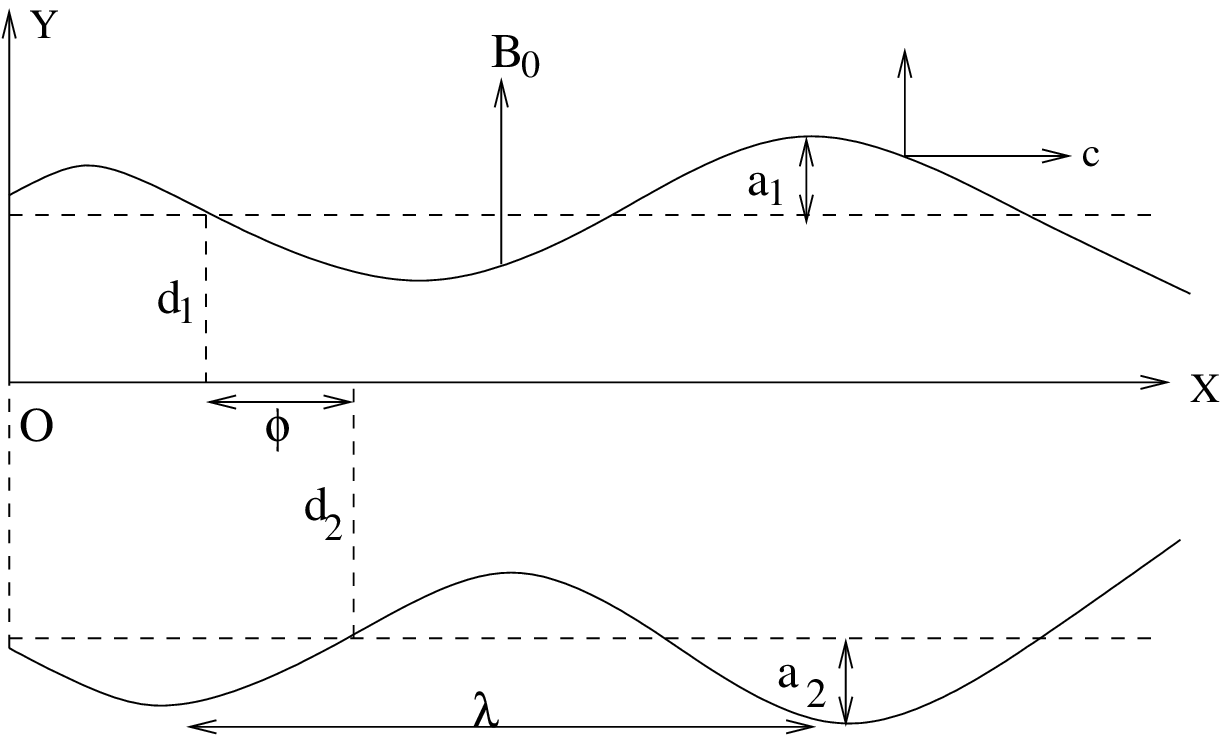}\\
 Fig. 1 ~~ A Physical sketch of the problem  \\
\end{center}
\end{minipage}\vspace*{.5cm}\\

\begin{minipage}{1.0\textwidth}
   \begin{center}
   \includegraphics[width=3.8in,height=2.5in ]{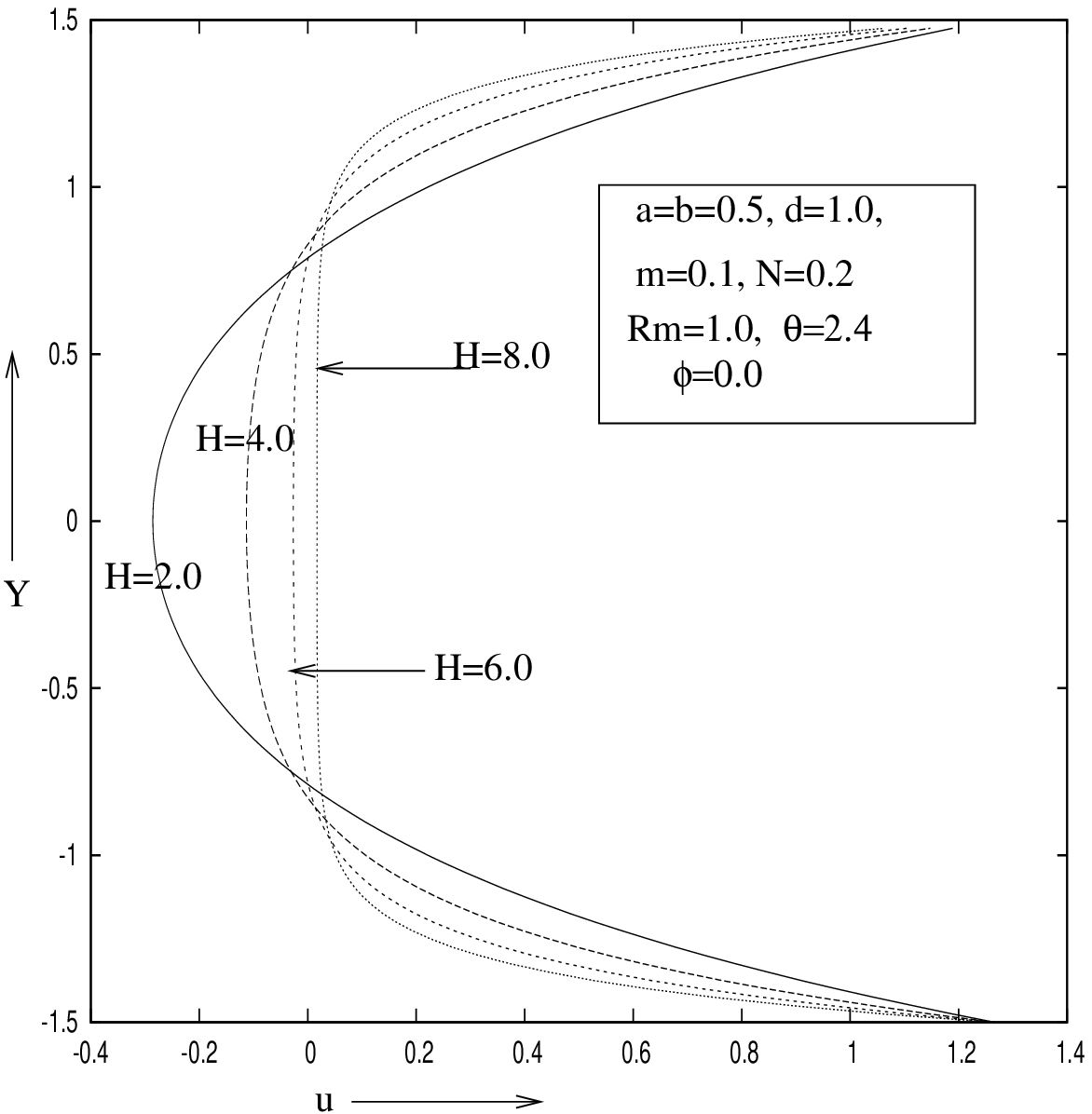}\\
 Fig. 2 Variation of axial velocity $u$ for different values of $H$ \\

\includegraphics[width=3.8in,height=2.5in ]{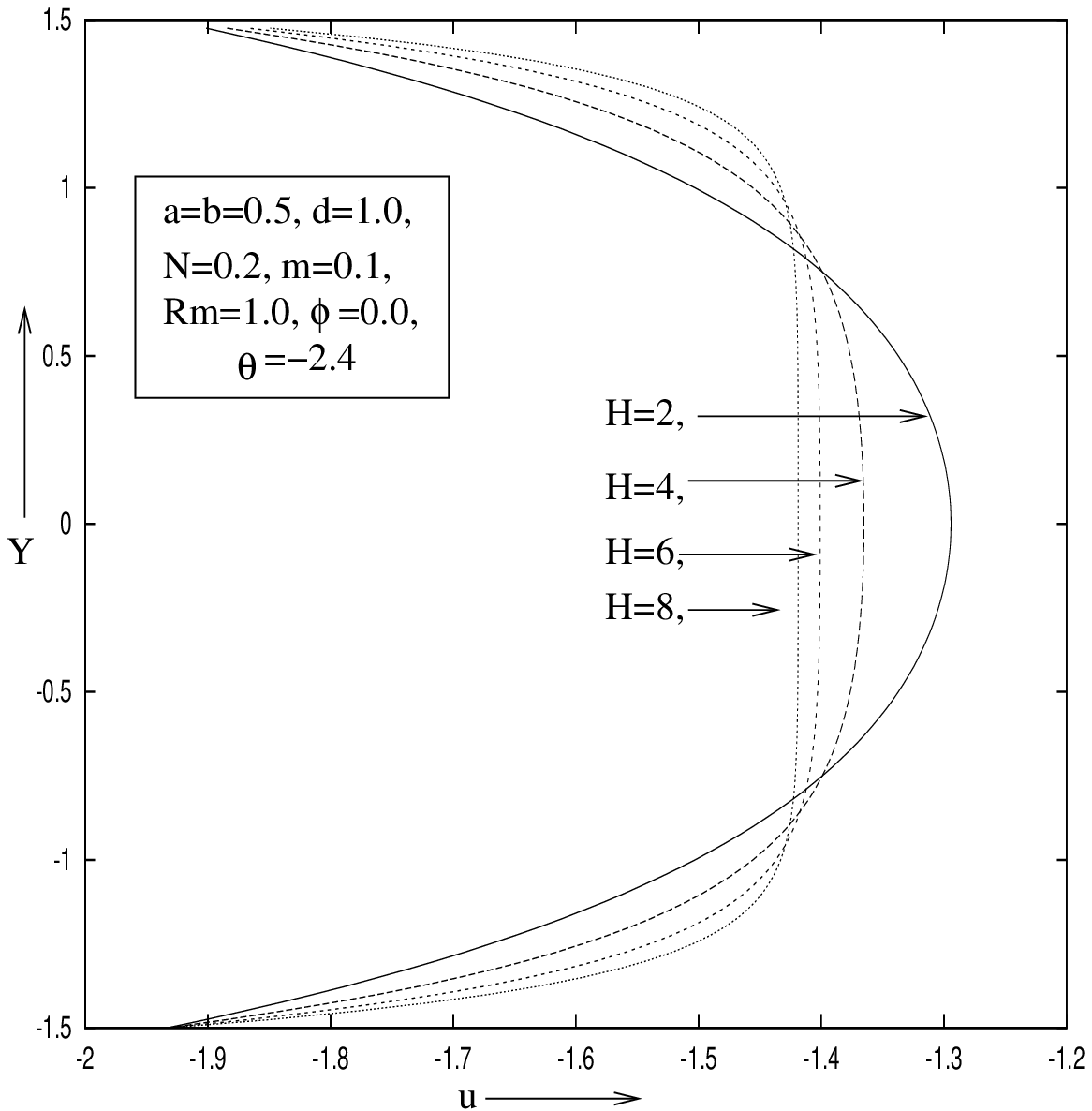}\\
Fig. 3 Variation of axial velocity $u$ for different values of
$H$\\
\includegraphics[width=3.8in,height=2.5in ]{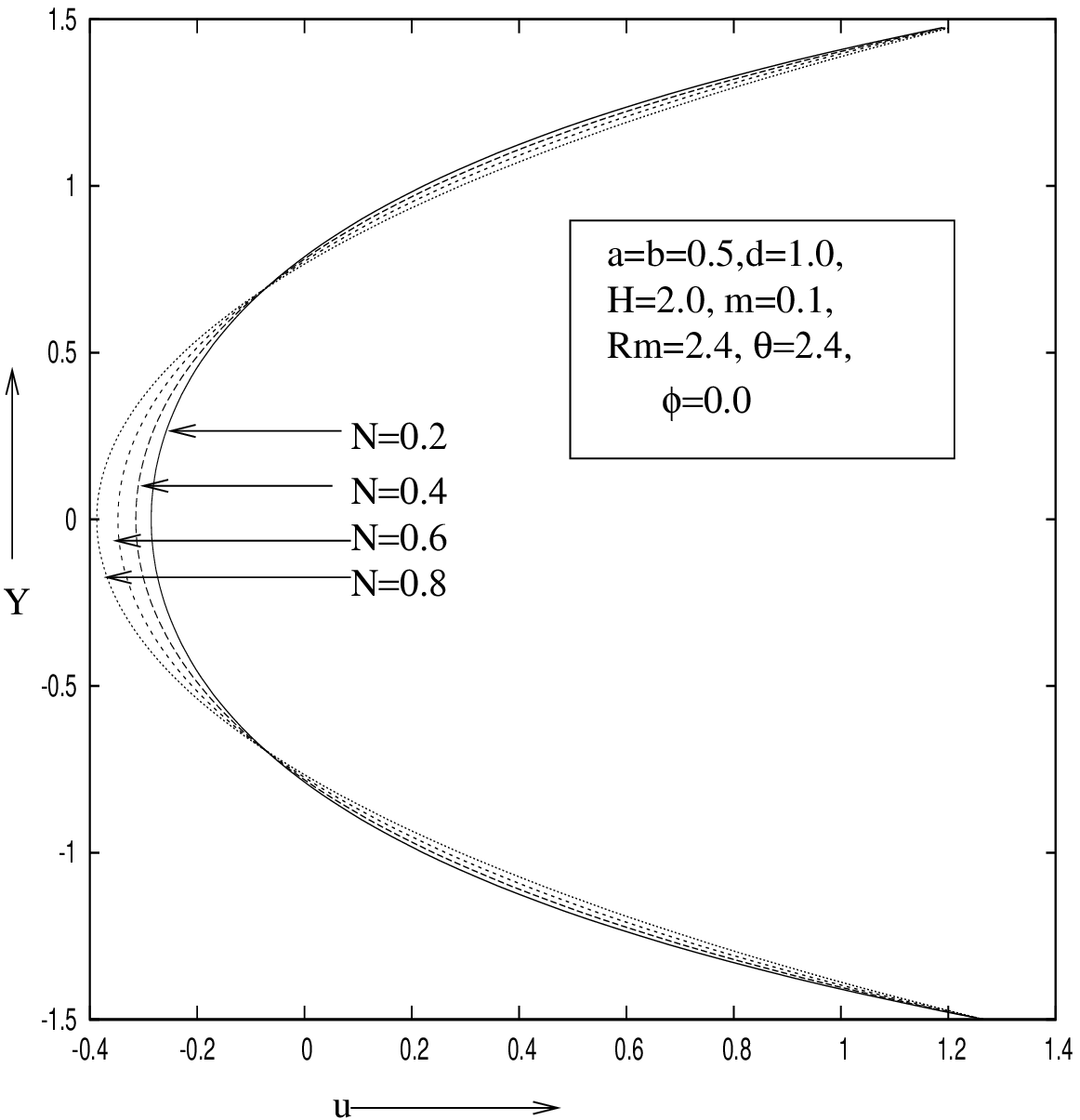}\\
Fig. 4 Variation of axial velocity $u$ for different values of
$N$\\
\end{center}
\end{minipage}\vspace*{.5cm}\\

\begin{minipage}{1.0\textwidth}
   \begin{center}
\includegraphics[width=3.8in,height=2.5in ]{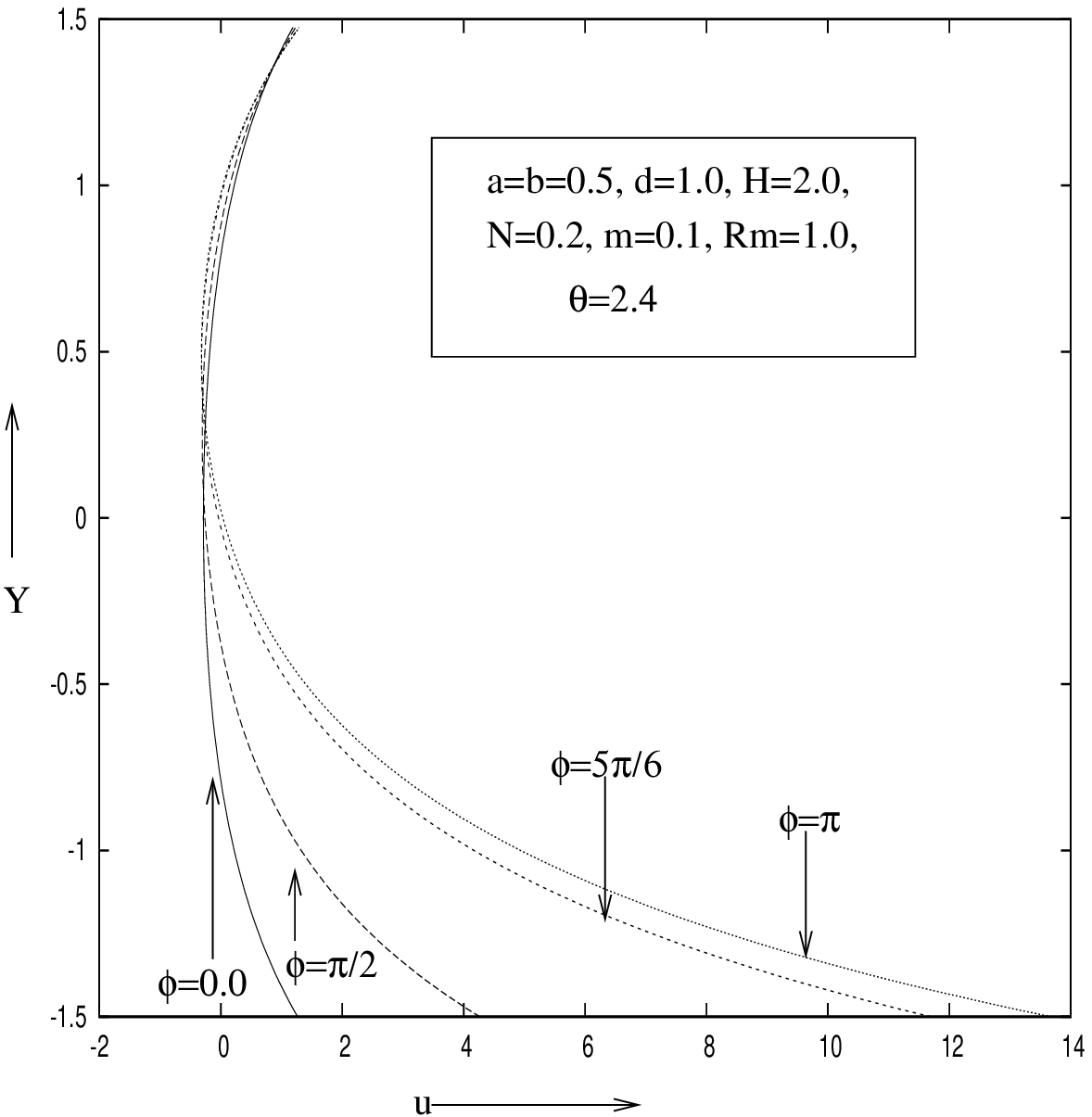}\\
Fig. 5 Variation of axial velocity $u$ for different values of $\phi$ \\

   \includegraphics[width=3.8in,height=2.5in ]{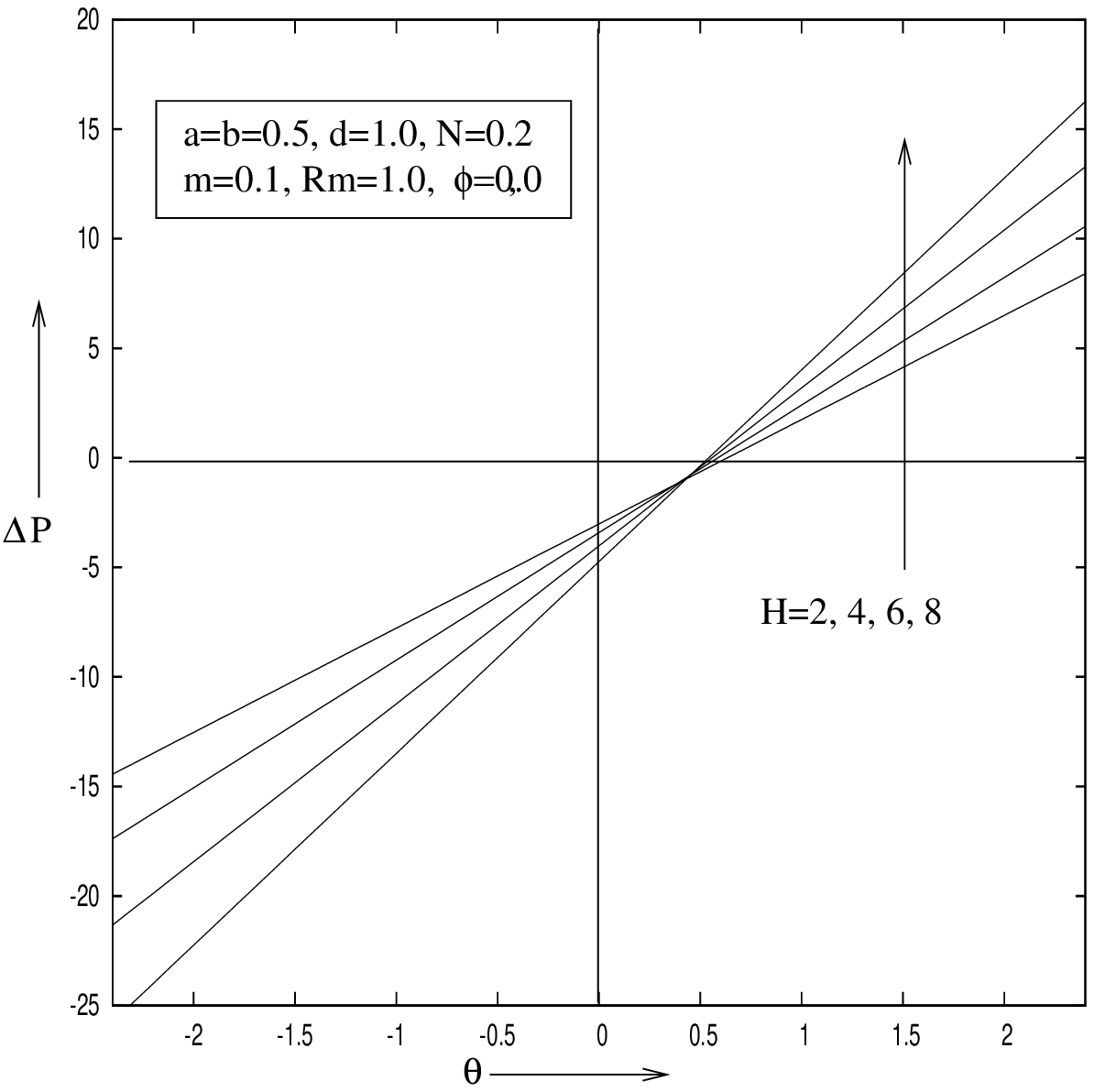}\\
Fig. 6 Variation of pressure rise $\triangle P$ with $\theta$ for different values of H\\

 \includegraphics[width=3.8in,height=2.5in ]{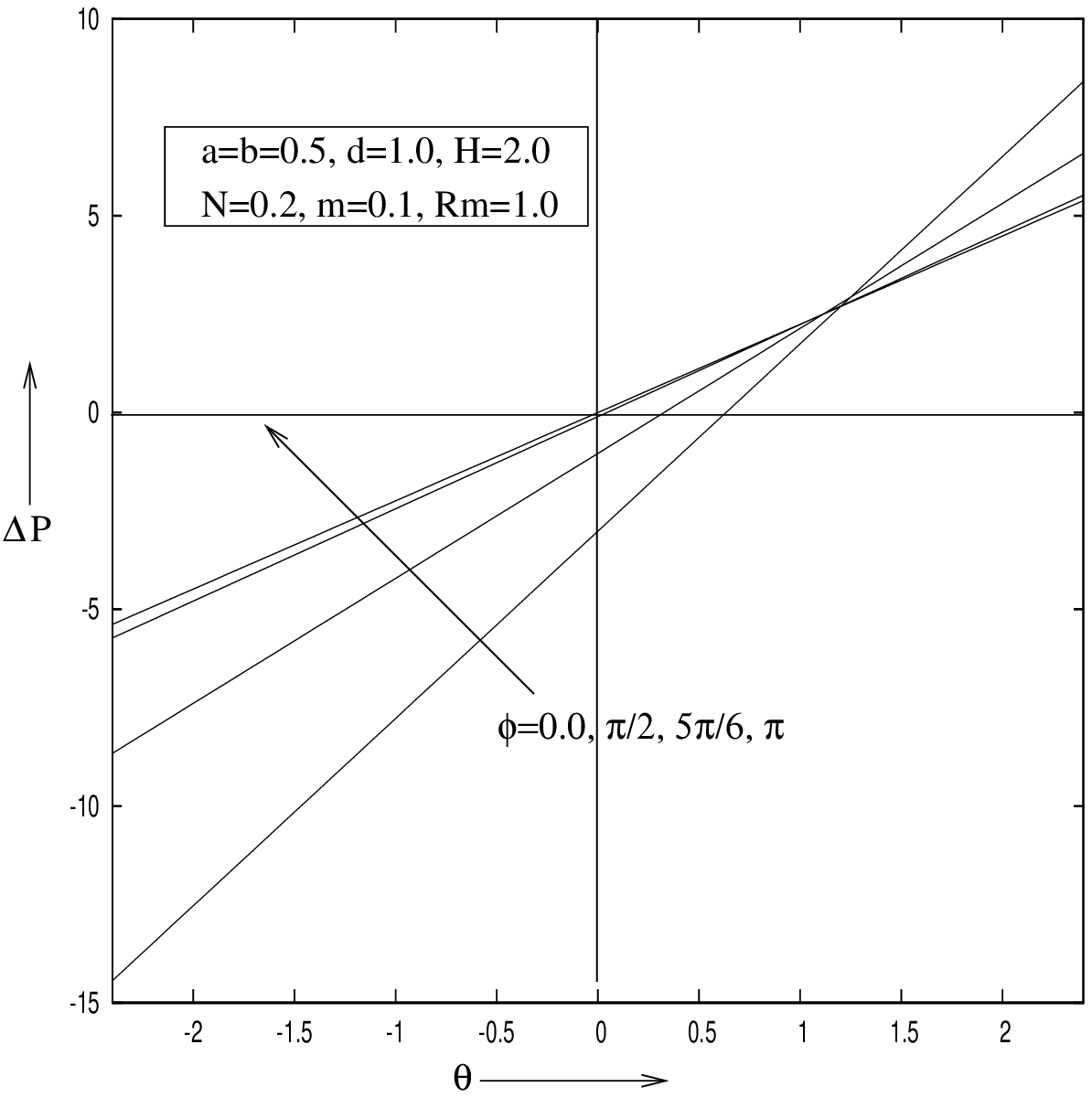}\\
 Fig. 7  Variation of pressure rise $\triangle P$ with $\theta$ for different values of $\phi$\\
\end{center}
\end{minipage}\vspace*{.5cm}\\

\begin{minipage}{1.0\textwidth}
   \begin{center}
\includegraphics[width=3.8in,height=2.5in ]{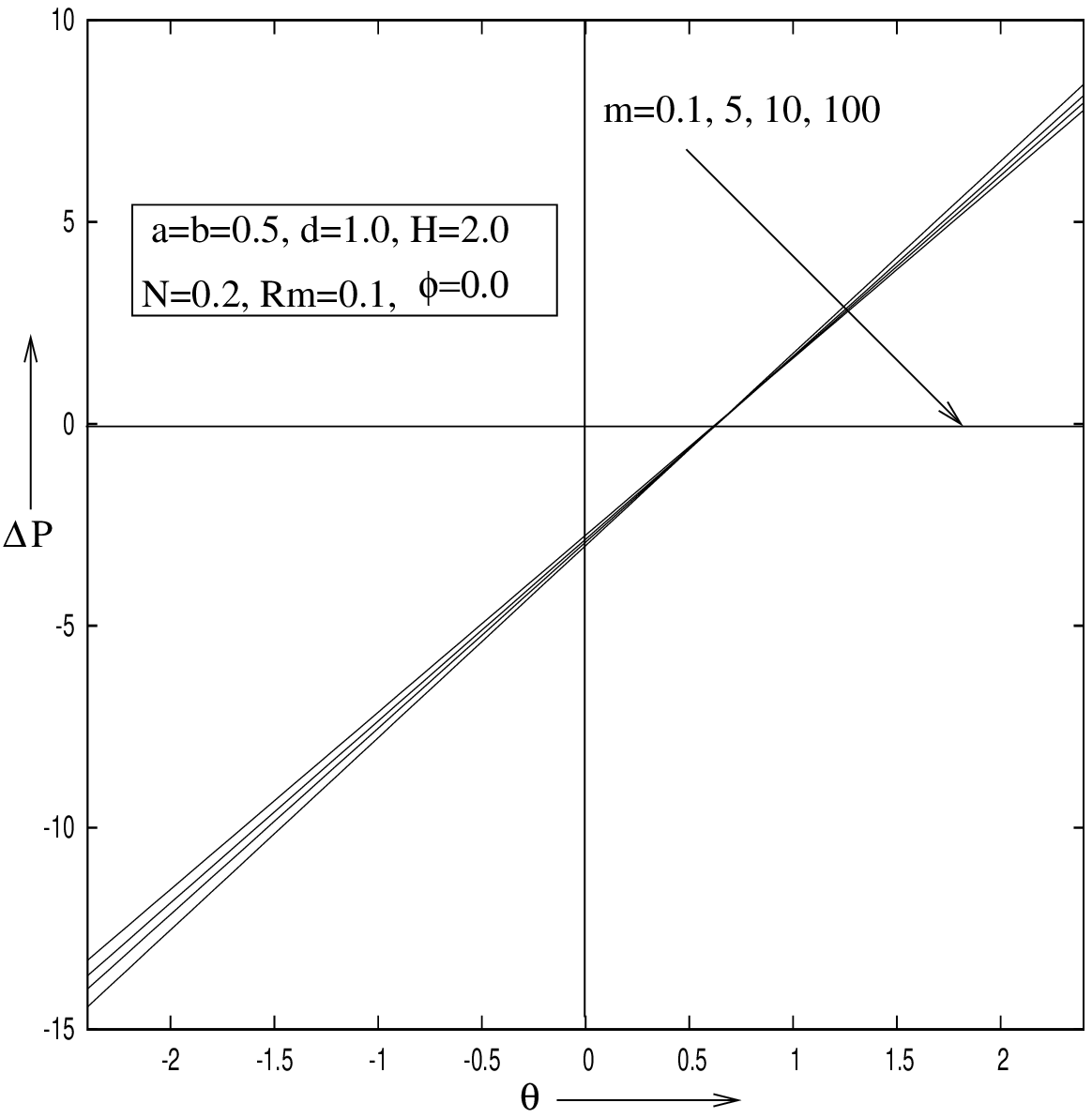}\\
Fig. 8~~ Variation of pressure rise $\triangle P$ with $\theta$ for different values of m\\

      \includegraphics[width=3.8in,height=2.5in ]{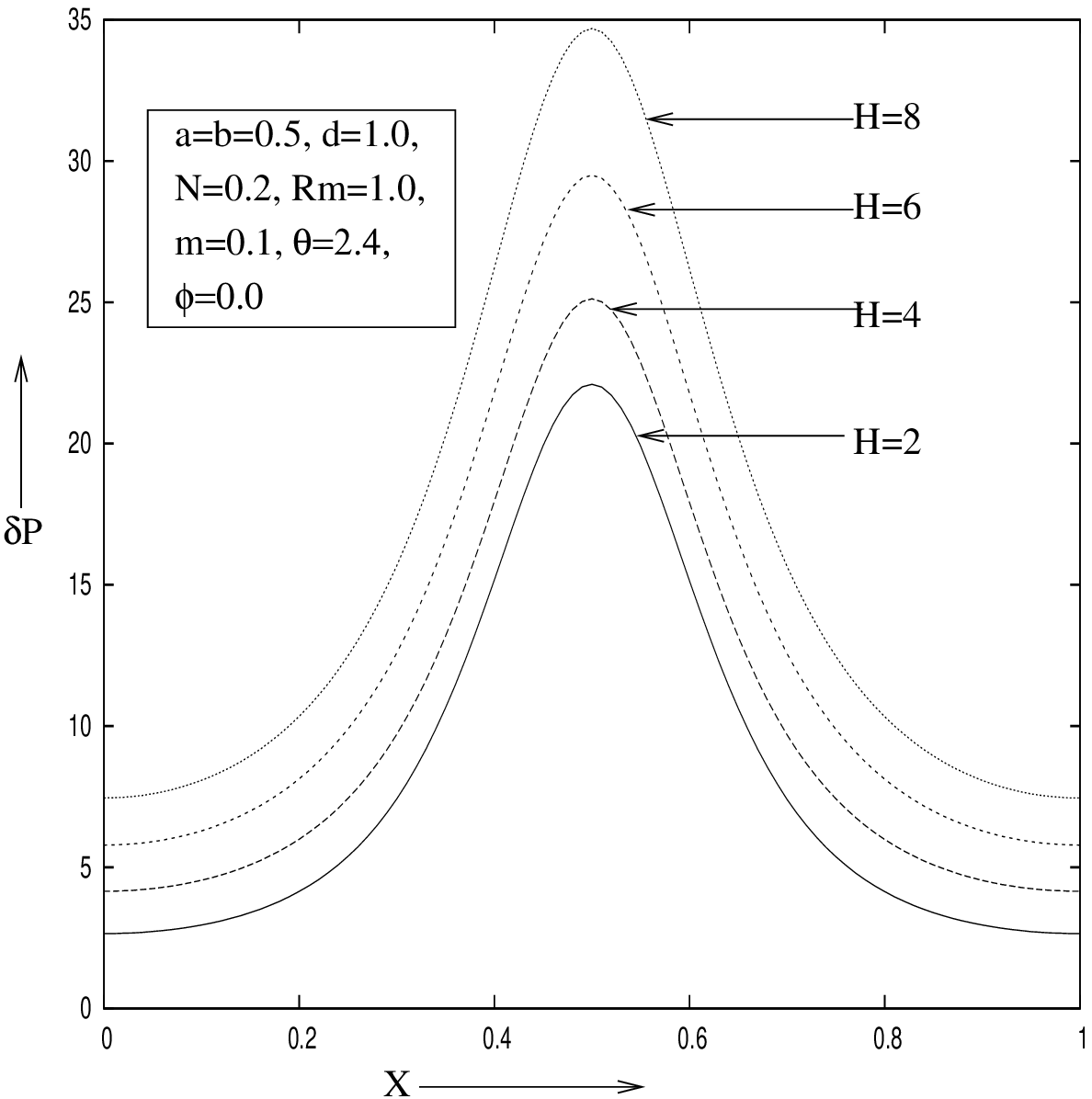}\\
Fig. 9 Distribution of pressure gradient $\delta
P=\frac{\partial p}{\partial x}$ for different values of ~$H$\\

\includegraphics[width=3.8in,height=2.5in ]{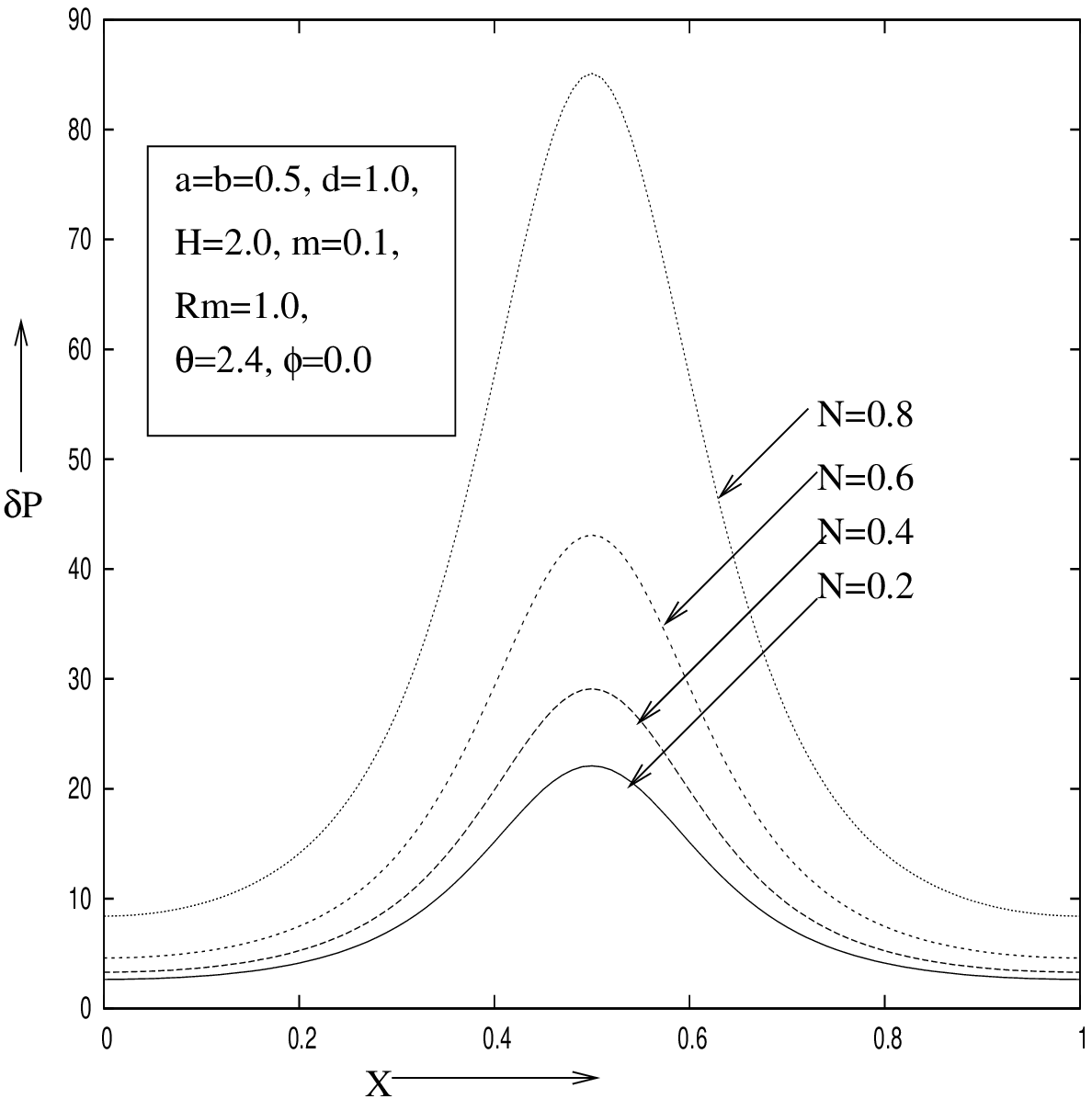}\\
Fig. 10 Distribution of pressure gradient $\delta
P=\frac{\partial p}{\partial x}$ for different values of ~$N$\\
\end{center}
\end{minipage}\vspace*{.5cm}\\

\begin{minipage}{1.0\textwidth}
   \begin{center}
\includegraphics[width=3.8in,height=2.5in ]{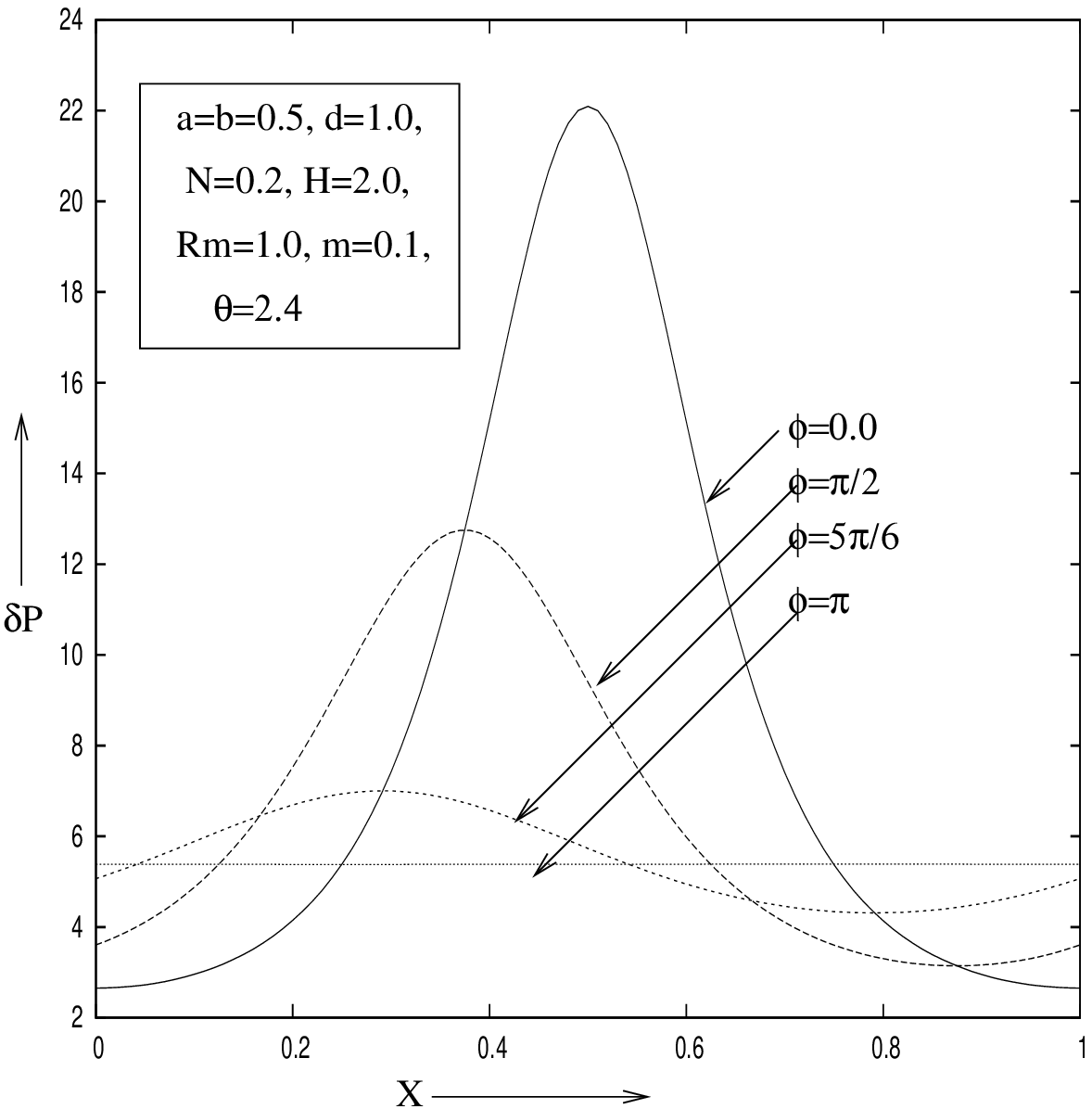}\\
Fig. 11 Distribution of pressure gradient $\delta
P=\frac{\partial p}{\partial x}$ for different values of ~$\phi$\\

 \includegraphics[width=3.8in,height=2.5in ]{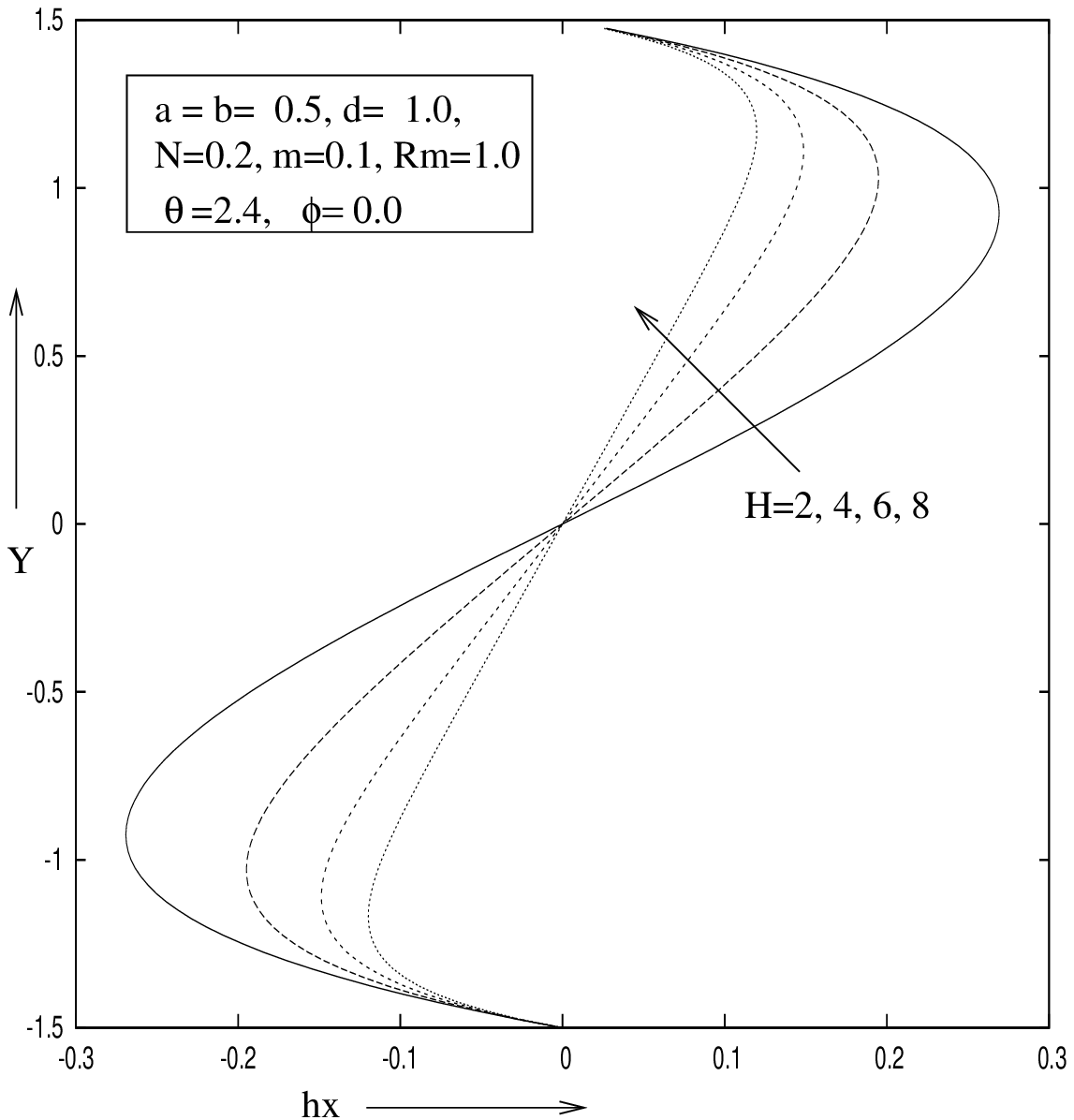}\\
Fig. 12 Effect of axial induced magnetic field $h_x$ for
different values of~$H$  \\

\includegraphics[width=3.8in,height=2.5in ]{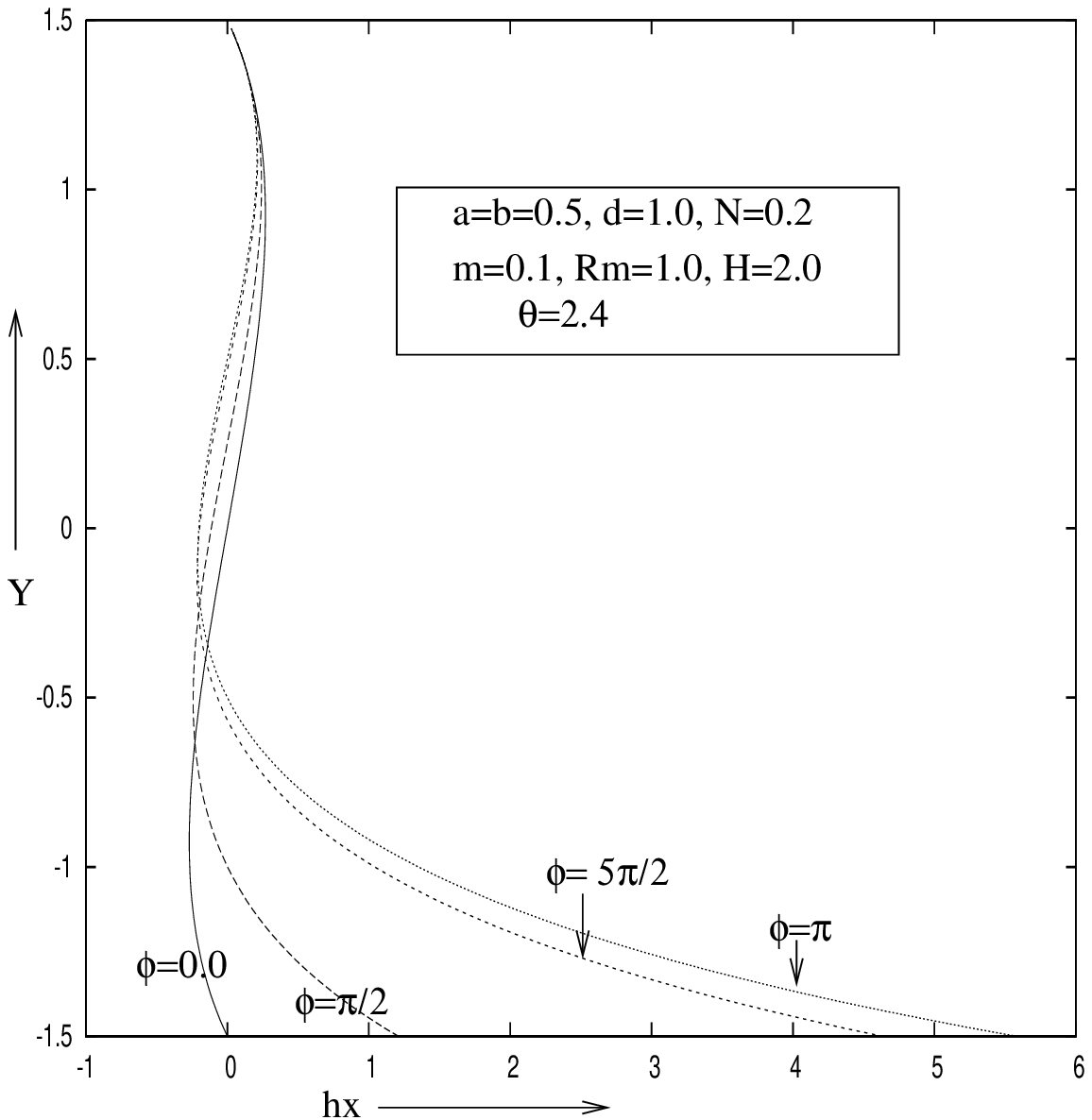}\\
Fig. 13  Effect of axial induced magnetic field $h_x$ for
different values of~$\phi$ \
    \end{center}
\end{minipage}\vspace*{.5cm}\\

\begin{minipage}{1.0\textwidth}
   \begin{center}
      \includegraphics[width=3.8in,height=2.5in ]{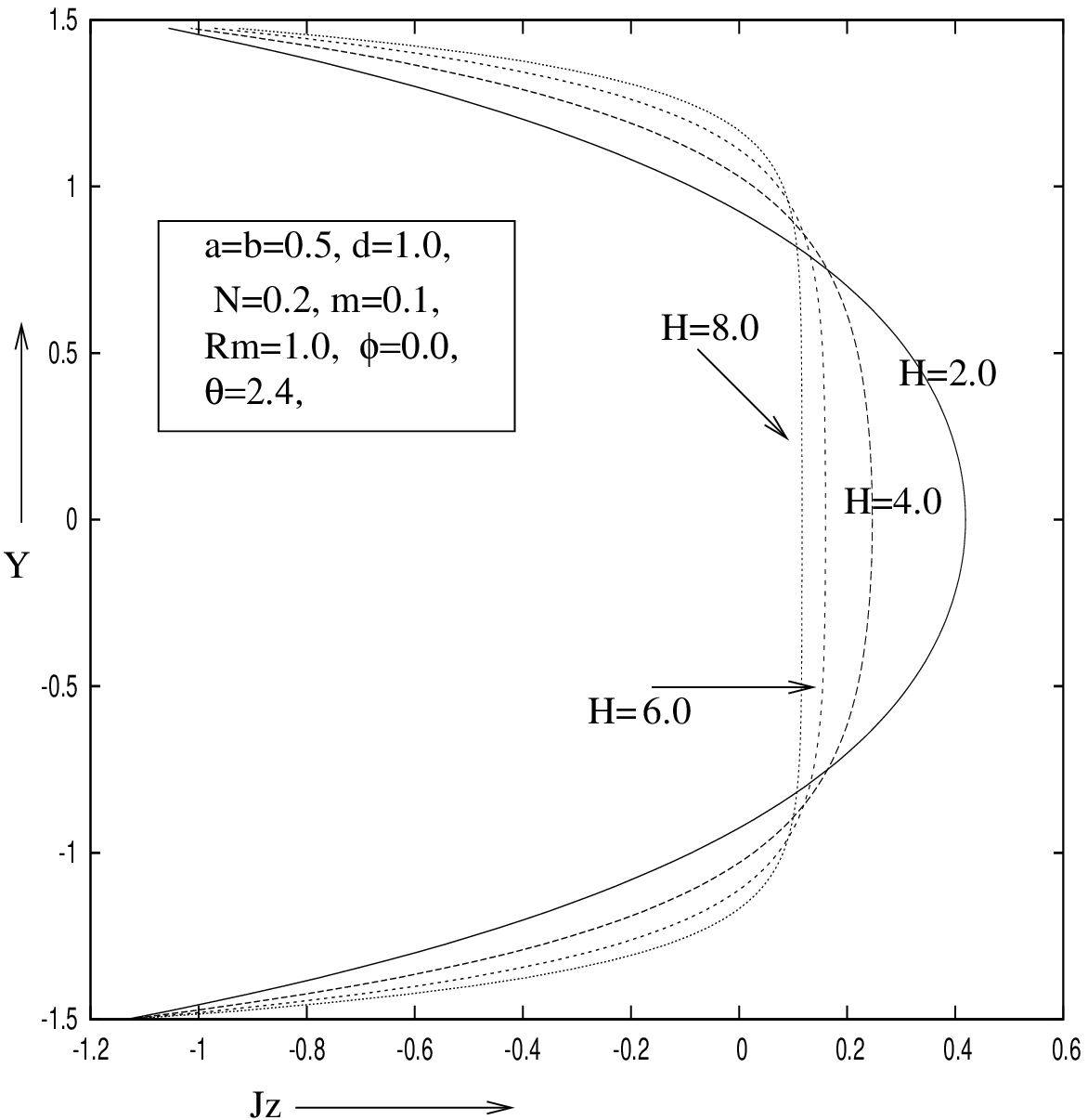}\\
Fig. 14 Distribution of current density $J_{z}$  for different values of~$H$ \\

\includegraphics[width=3.8in,height=2.5in ]{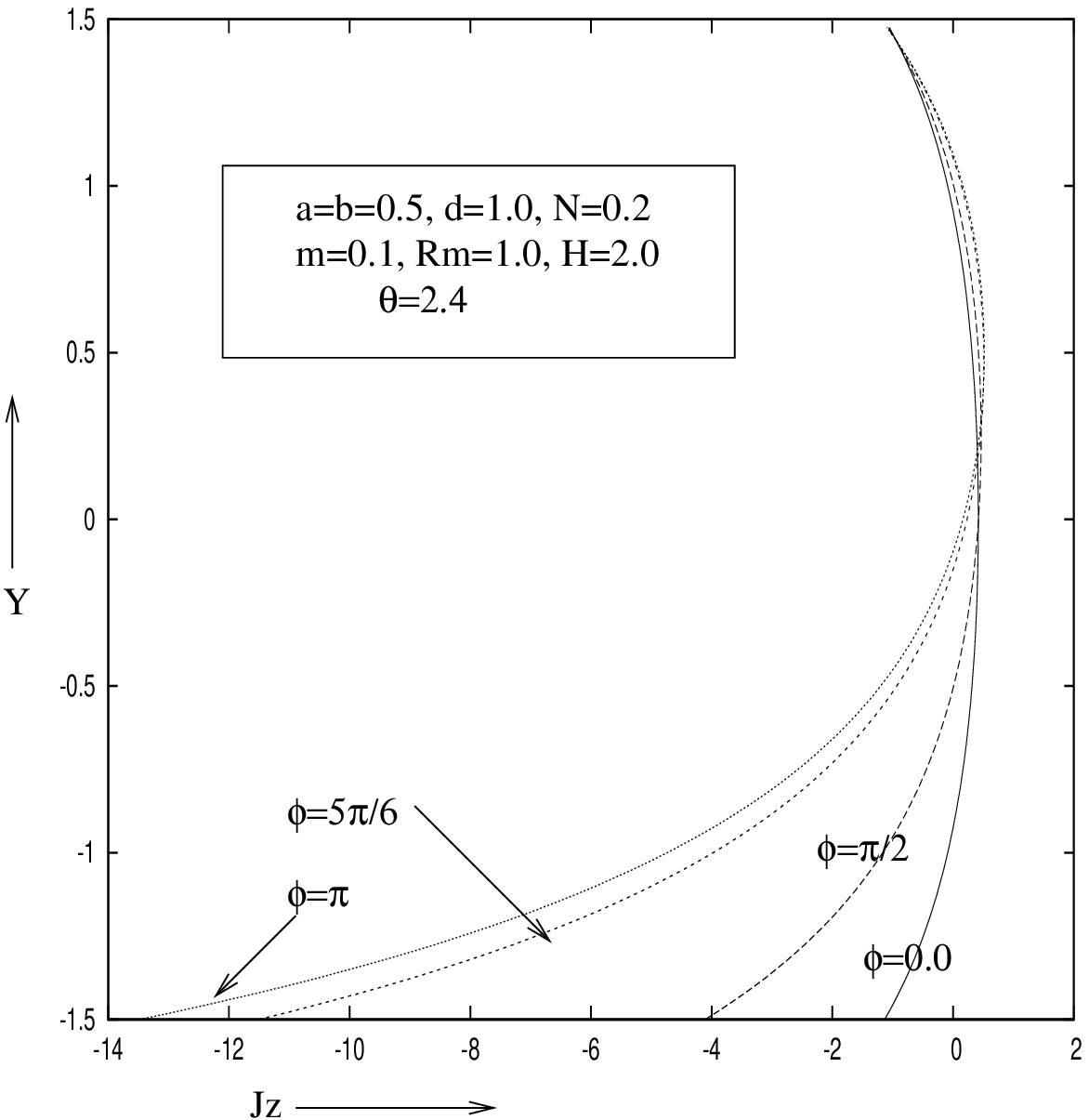}\\
Fig. 15  Distribution of current density $J_{z}$ for different
values of ~$\phi$\\

      \includegraphics[width=3.8in,height=2.5in ]{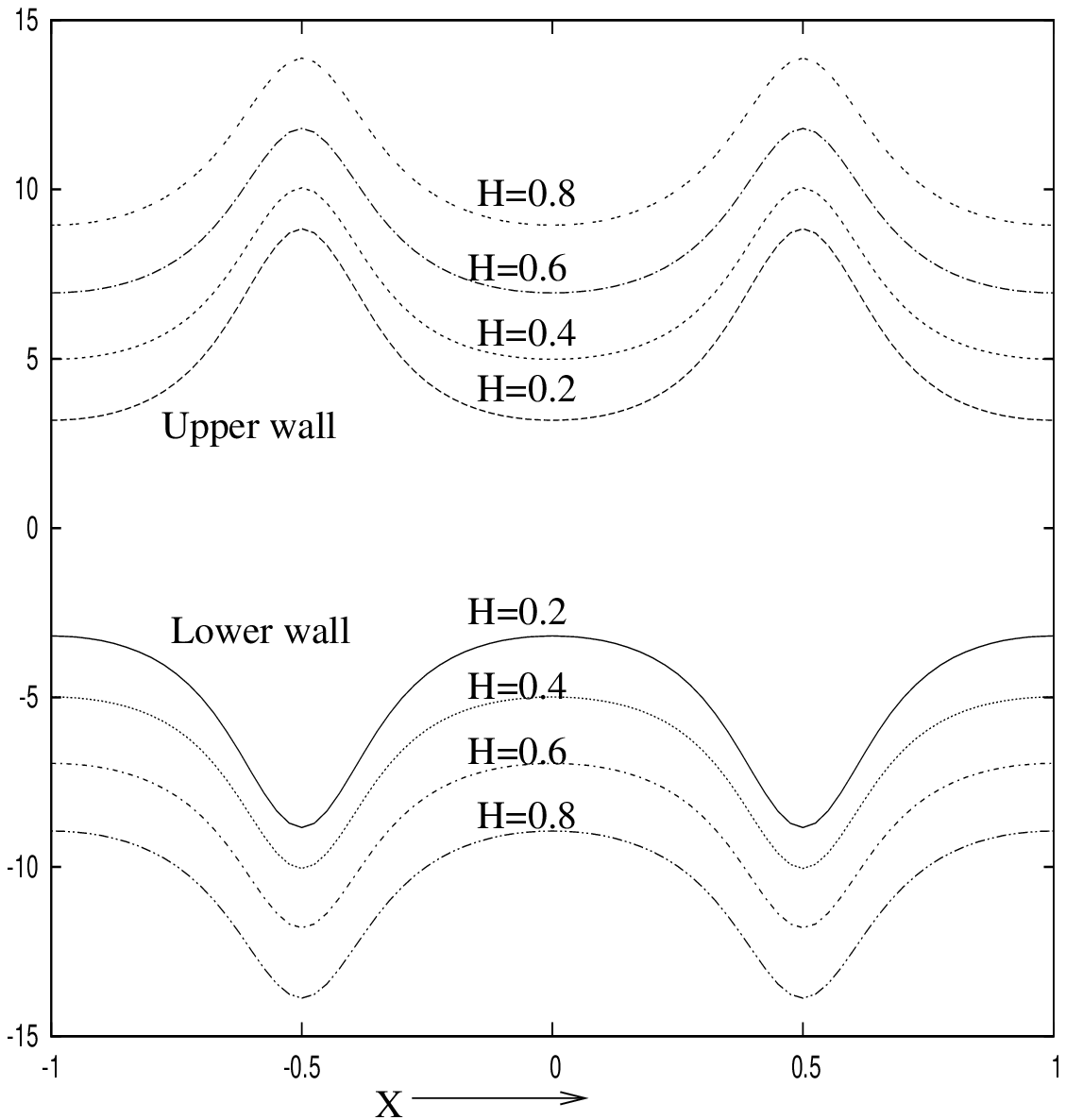}\\
Fig. 16  Variation of $\tau_{xy}$ along with the co-ordinate $x$
for different values of $H$ with $a=b=0.5$, $d=1.0$, $N=0.2$, $m=0.1$, $R_{m}=1.0$, $\theta=2.4$, $\phi=0.0$\\
   \end{center}
\end{minipage}\vspace*{.5cm}\\

\begin{minipage}{1.0\textwidth}
   \begin{center}
      \includegraphics[width=3.8in,height=2.5in ]{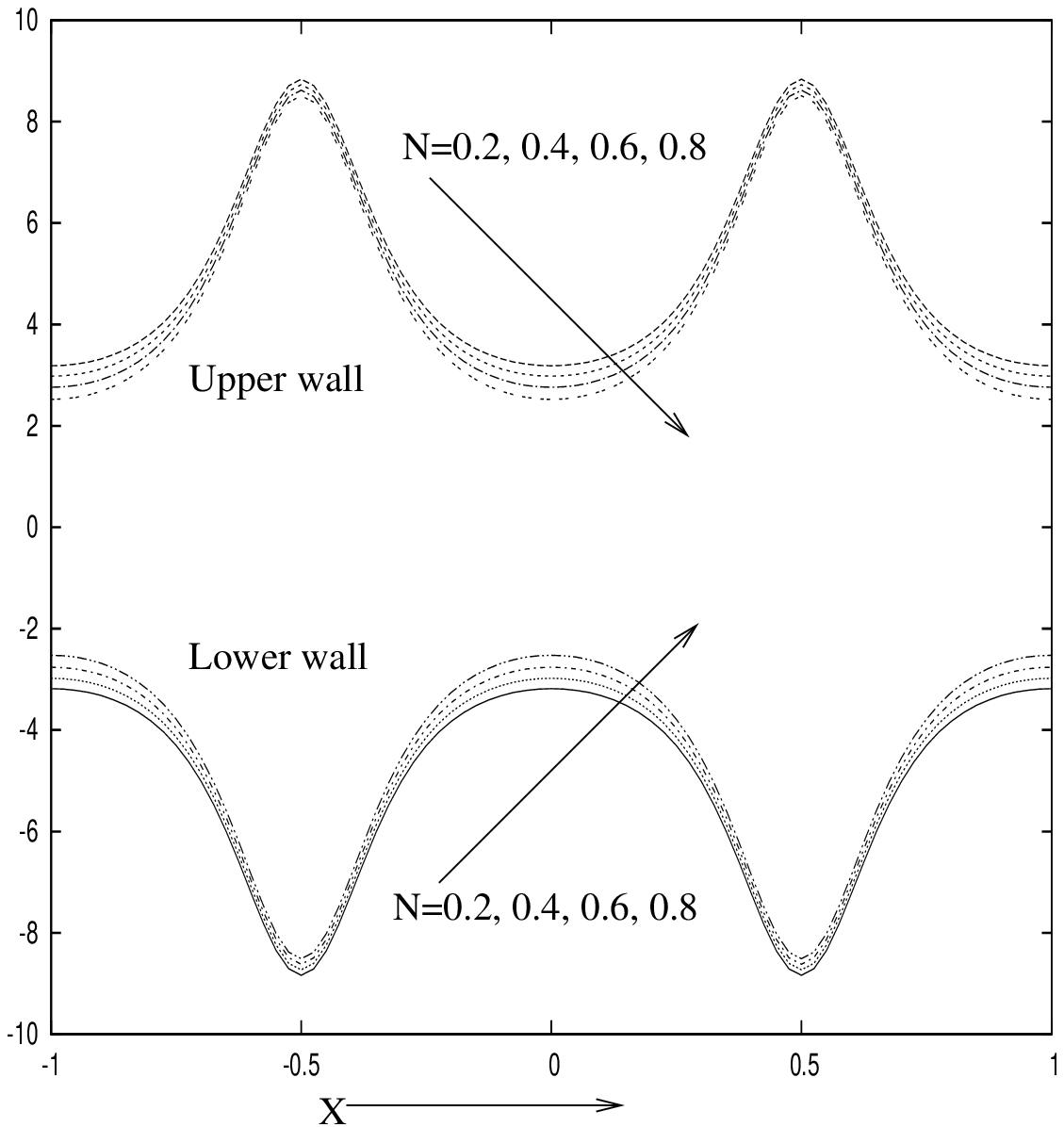}\\
Fig. 17 Variation of $\tau_{xy}$ along with the co-ordinate $x$
for different values of $N$ with $a=b=0.5$, $d=1.0$, $H=2.0$, $m=0.1$, $R_{m}=1.0$, $\theta=2.4$, $\phi=0.0$\\

   \includegraphics[width=3.8in,height=2.5in ]{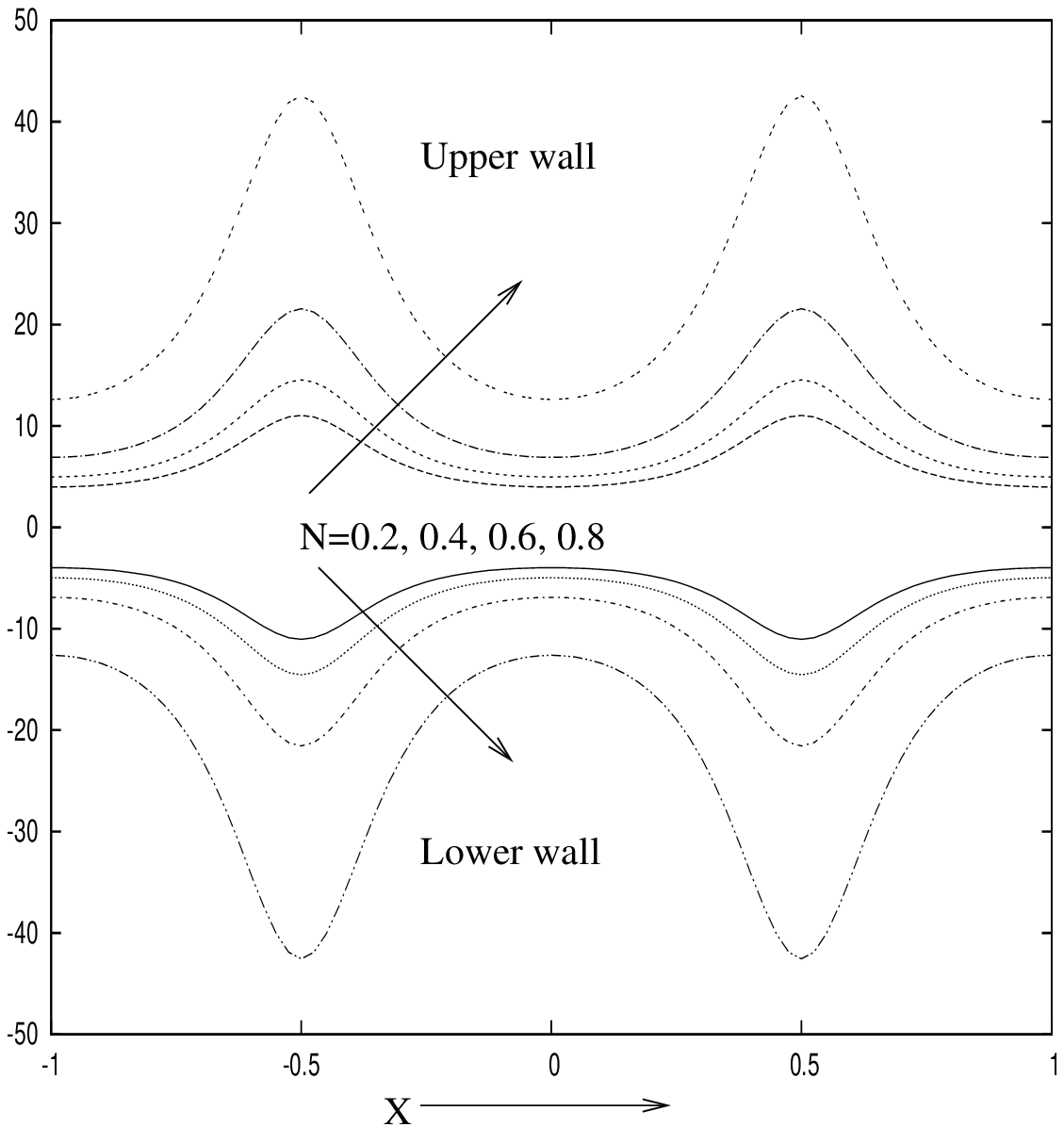}\\
Fig. 18  Variation of $\tau_{yx}$ along with the co-ordinate $x$
for different values of $N$ with $a=b=0.5$, $d=1.0$, $H=2.0$,
$m=0.1$, $R_{m}=1.0$, $\theta=2.4$, $\phi=0.0$\\

\includegraphics[width=3.8in,height=2.5in ]{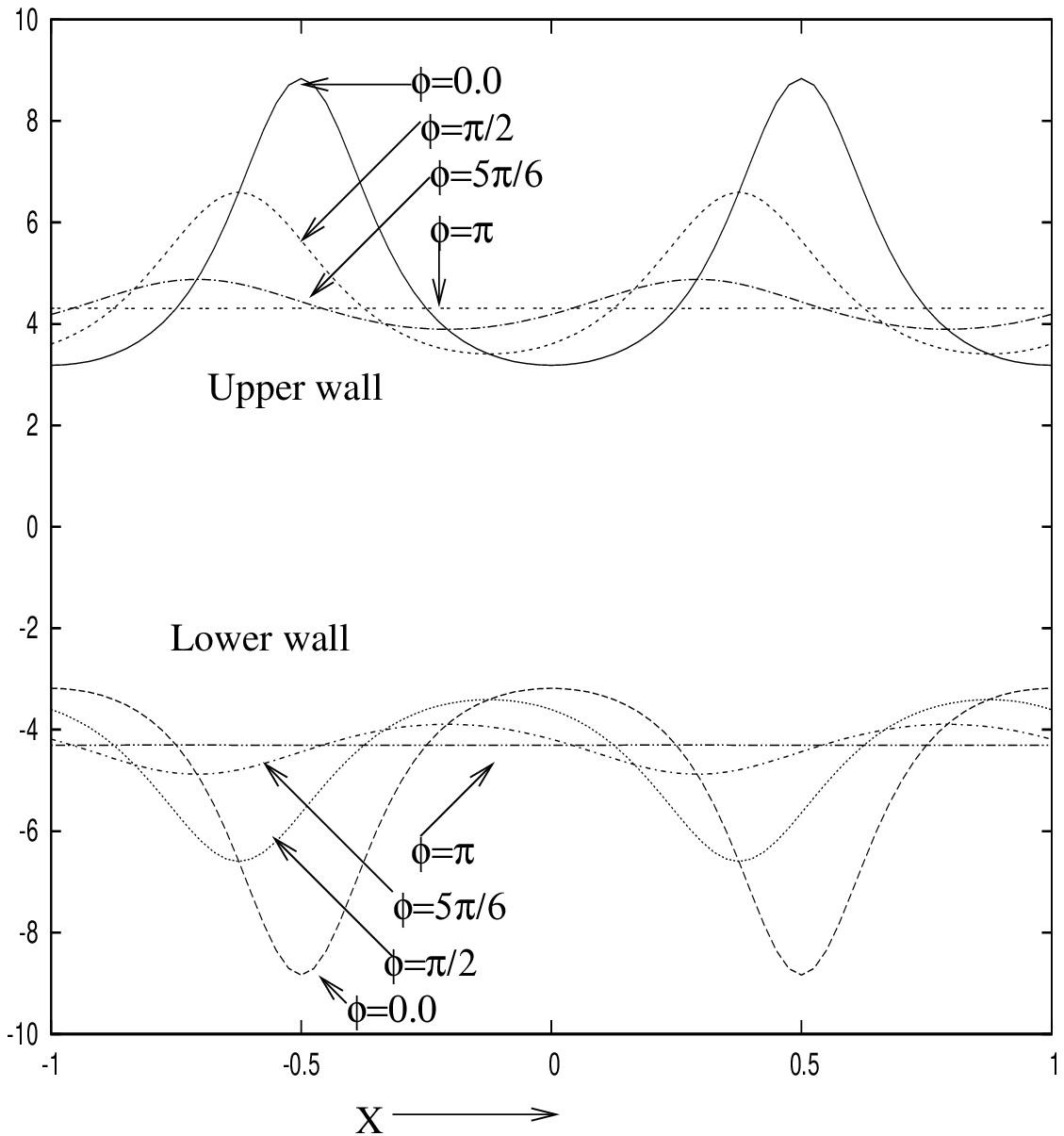}\\
Fig. 19 Variation of $\tau_{xy}$ along with the co-ordinate $x$
for different values of $\phi$ with $a=b=0.5$, $d=1.0$, $N=0.2$, $m=0.1$, $R_{m}=1.0$, $\theta=2.4$, $H=2.0$\\
  \end{center}
\end{minipage}\vspace*{.5cm}\\

\begin{minipage}{1.0\textwidth}
   \begin{center}
\includegraphics[width=3.8in,height=2.4in ]{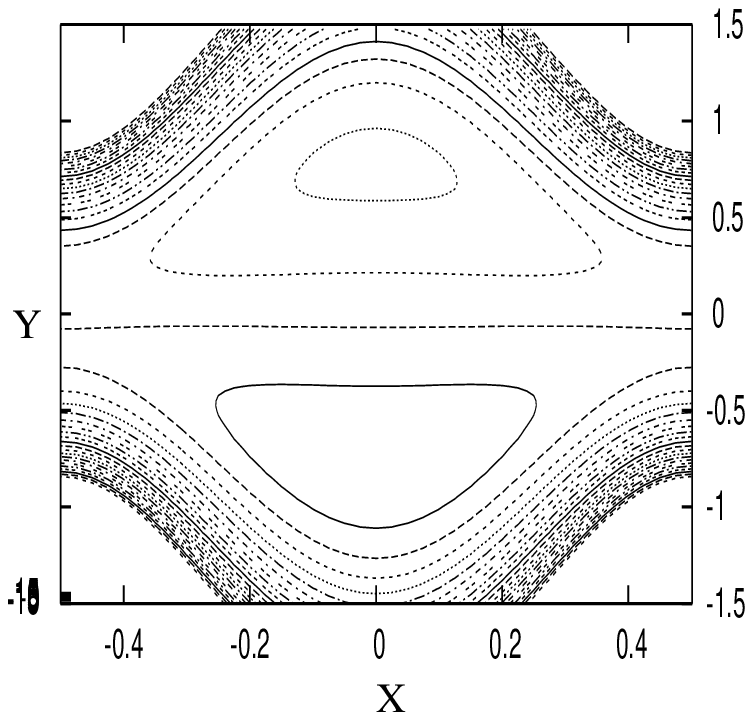}\\
 Fig. 20  Streamlines for $H=2$ with $a=b=0.5$, $d=1.0$, $\phi=0.0$, $N=0.2$, $m=0.1$, $\theta=2.4$, $R_{m}=1.0$ \\

\includegraphics[width=3.8in,height=2.4in ]{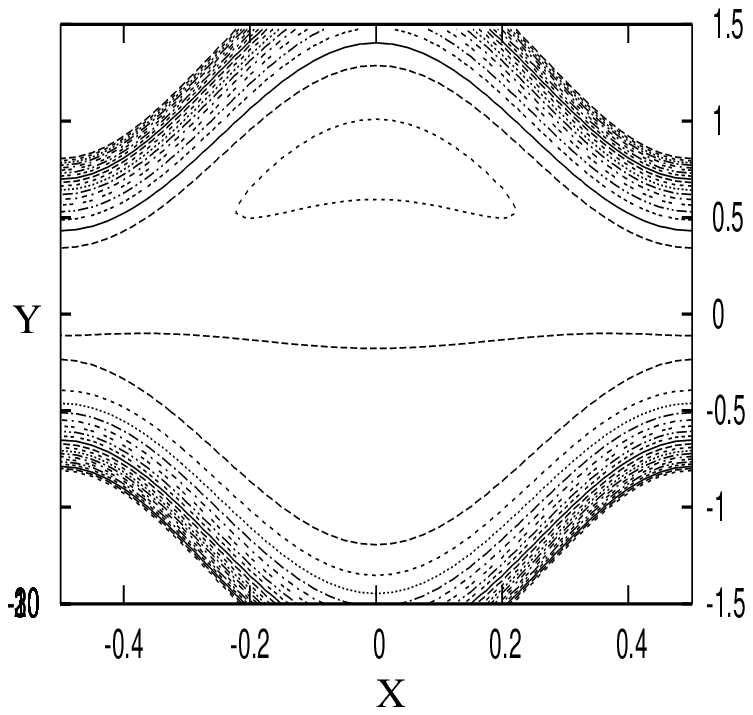}\\
Fig. 21 Streamlines for $H=4$ with $a=b=0.5$, $d=1.0$, $\phi=0.0$, $N=0.2$, $m=0.1$, $\theta=2.4$, $R_{m}=1.0$ \\

\includegraphics[width=3.8in,height=2.4in ]{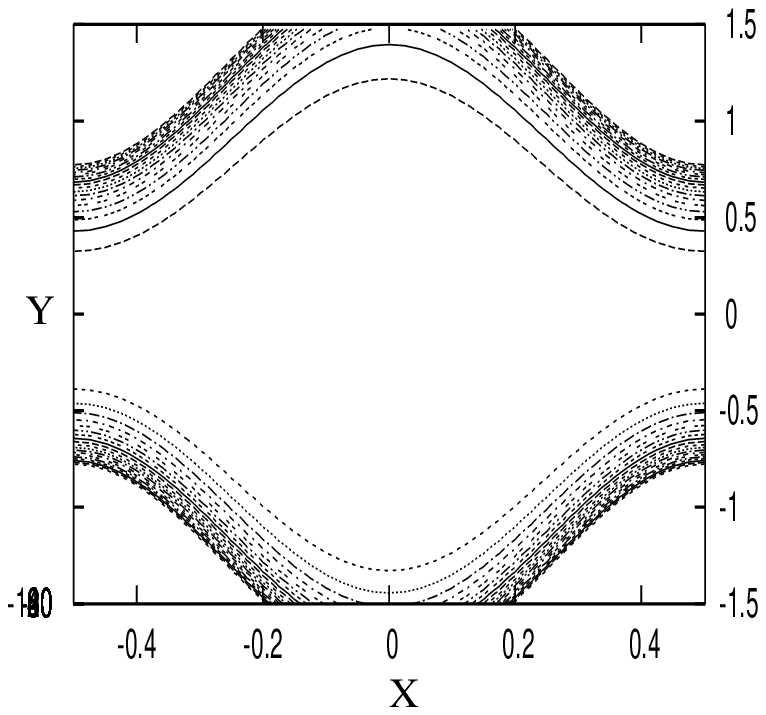}\\
 Fig. 22  Streamlines for $H=6$ with $a=b=0.5$, $d=1.0$, $\phi=0.0$, $N=0.2$, $m=0.1$, $\theta=2.4$, $R_{m}=1.0$ \\
\end{center}
\end{minipage}\vspace*{.5cm}\\

\begin{minipage}{1.0\textwidth}
   \begin{center}
\includegraphics[width=3.8in,height=2.4in ]{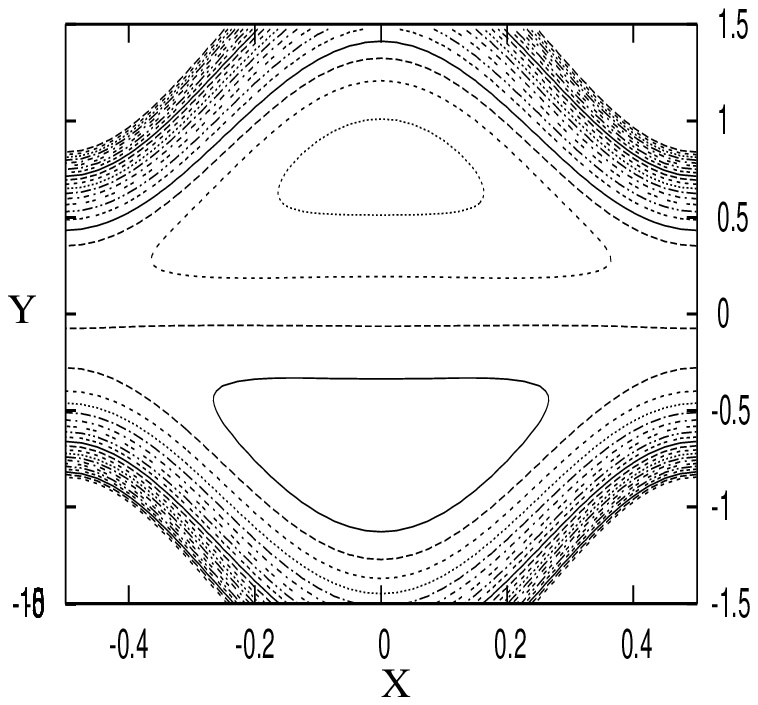}\\
Fig. 23 Streamlines for $N=0.4$ with $a=b=0.5$, $d=1.0$, $\phi=0.0$, $H=2.0$, $m=0.1$, $\theta=2.4$, $R_{m}=1.0$ \\

\includegraphics[width=3.8in,height=2.4in ]{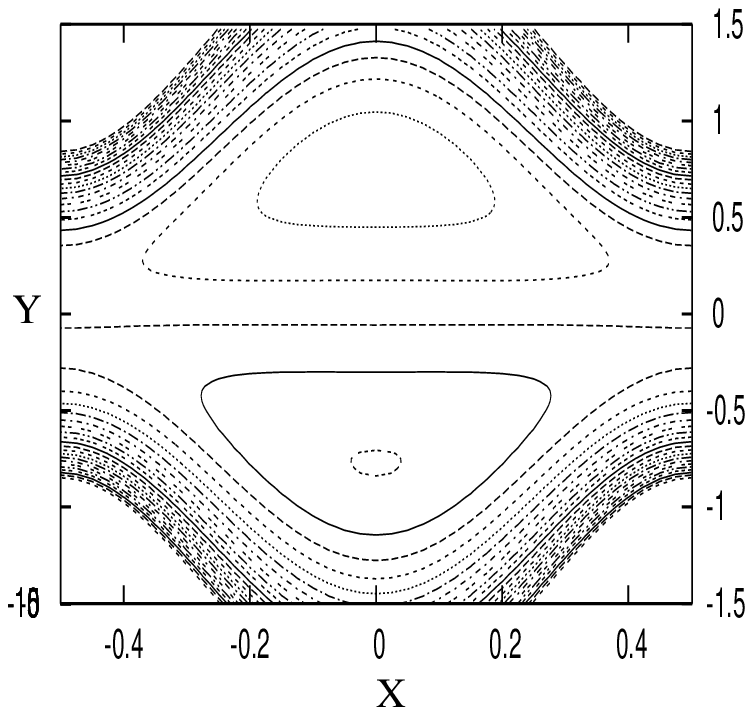}\\
 Fig. 24 Streamlines for $N=0.6$ with $a=b=0.5$, $d=1.0$, $\phi=0.0$, $H=2.0$, $m=0.1$, $\theta=2.4$, $R_{m}=1.0$ \\

\includegraphics[width=3.8in,height=2.4in ]{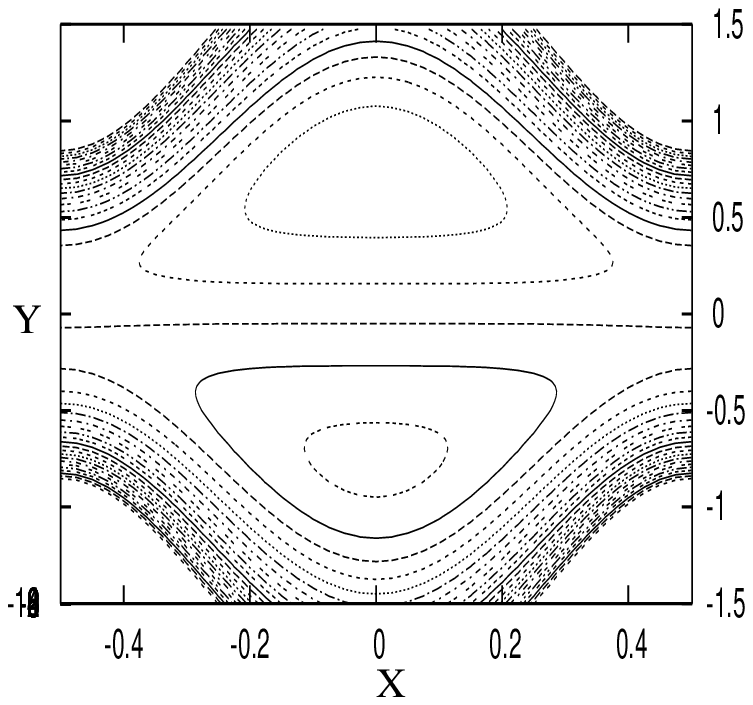}\\
Fig. 25 Streamlines for $N=0.8$ with $a=b=0.5$, $d=1.0$, $\phi=0.0$, $H=2.0$, $m=0.1$, $\theta=2.4$, $R_{m}=1.0$ \\
\end{center}
\end{minipage}\vspace*{.5cm}\\

\begin{minipage}{1.0\textwidth}
   \begin{center}
\includegraphics[width=3.8in,height=2.4in ]{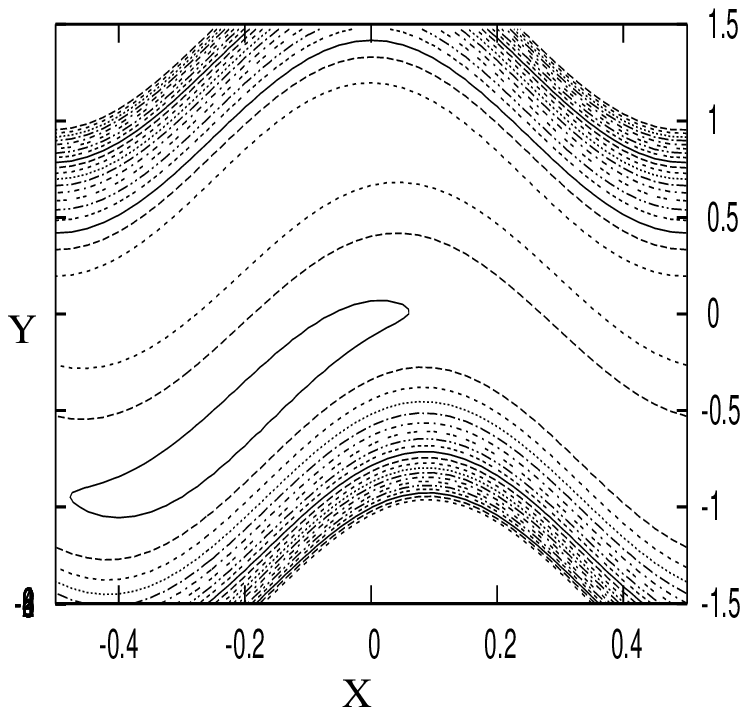}\\
Fig. 26 Streamlines for $\phi=\pi/2$ with $a=b=0.5$, $d=1.0$, $m=0.1$, $H=2.0$, $N=0.2$, $\theta=2.4$, $R_{m}=1.0$ \\

   \includegraphics[width=3.8in,height=2.4in ]{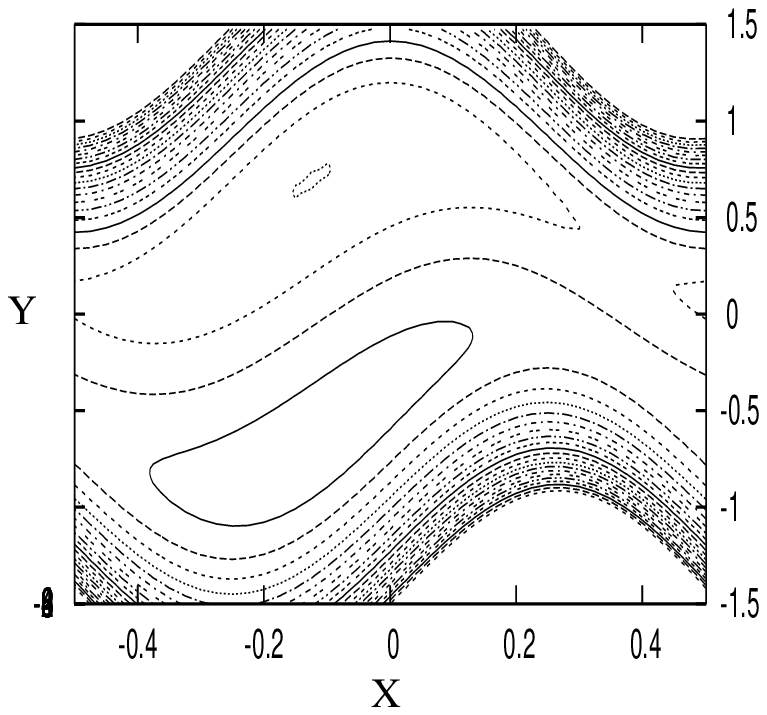}\\
 Fig. 27 Streamlines for $\phi=5\pi/6$ with $a=b=0.5$, $d=1.0$, $m=0.1$, $H=2.0$, $N=0.2$, $\theta=2.4$, $R_{m}=1.0$ \\

\includegraphics[width=3.8in,height=2.4in ]{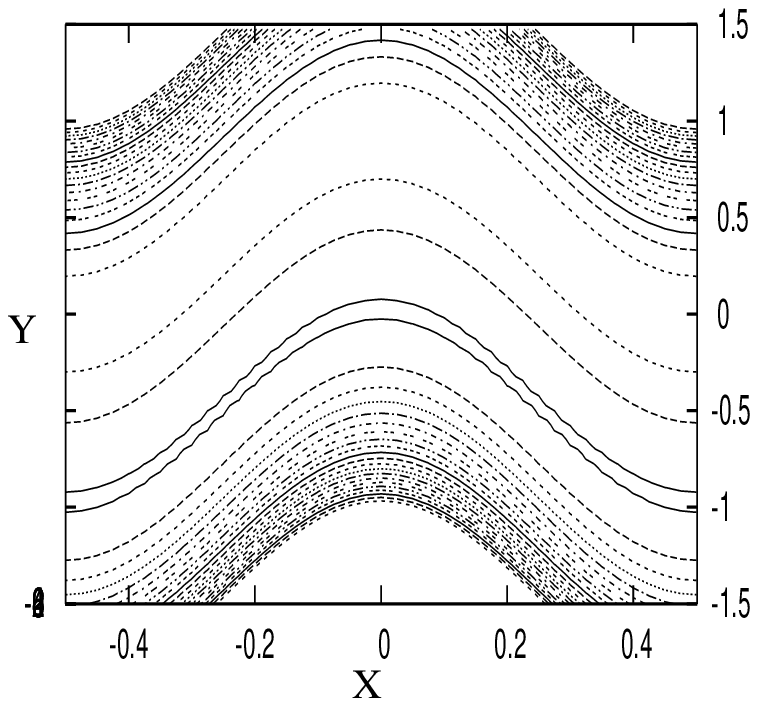}\\
Fig. 28 Streamlines for $\phi=\pi$ with $a=b=0.5$, $d=1.0$, $m=0.1$, $H=2.0$, $N=0.2$, $\theta=2.4$, $R_{m}=1.0$ \\
\end{center}
\end{minipage}\vspace*{.5cm}\\

\begin{minipage}{1.0\textwidth}
   \begin{center}
   \includegraphics[width=3.8in,height=2.4in ]{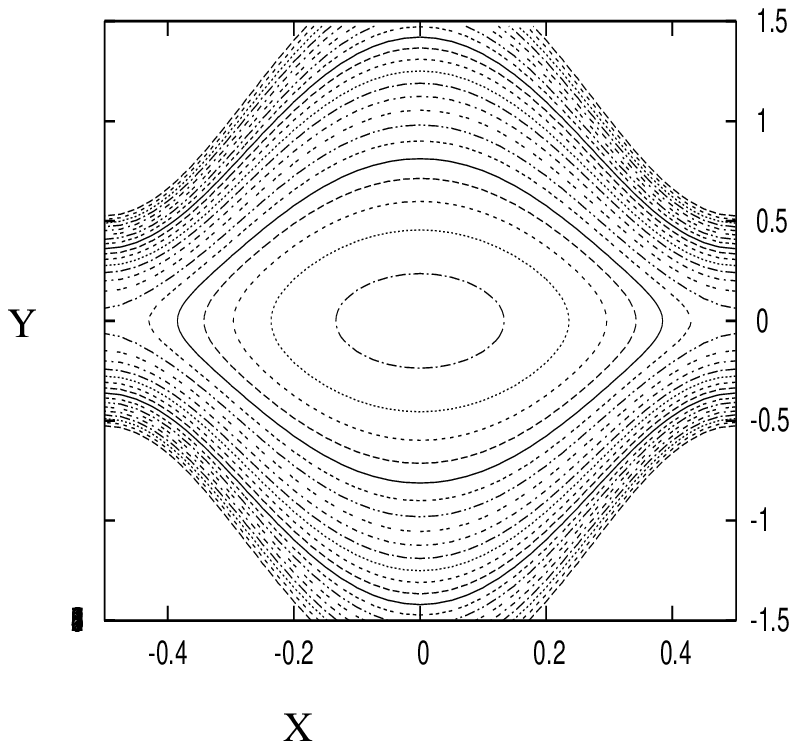}\\
 Fig. 29  Distribution of magnetic force function $\phi(x,y)$ for $H=2$\\
                 with $a=b=0.5$, $d=1.0$, $N=0.2$, $m=0.1$, $\phi=0.0$, $\theta=2.4$, $R_{m}=1.0$\\
\includegraphics[width=3.8in,height=2.4in ]{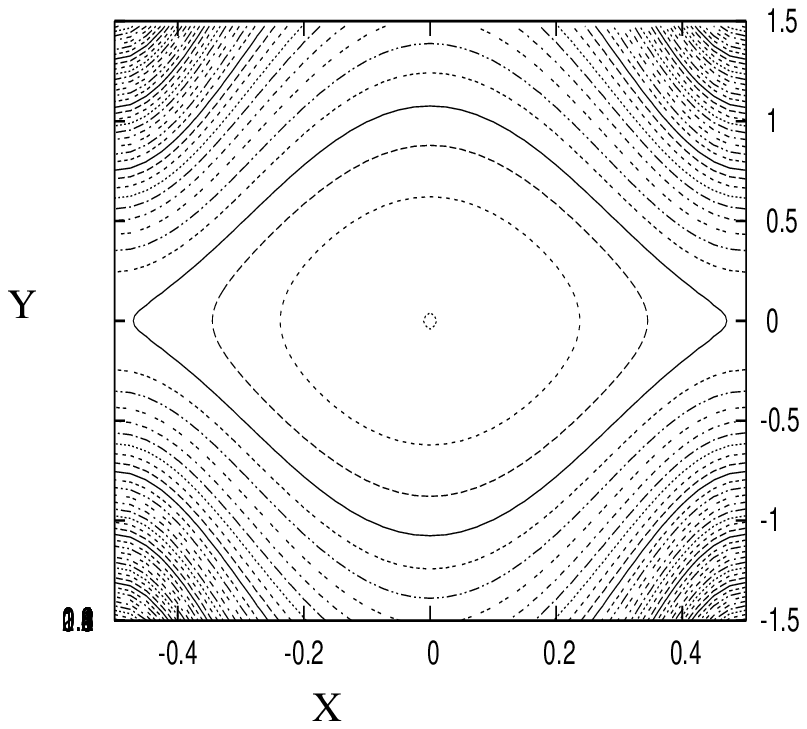}\\
Fig. 30 Distribution of magnetic force function $\phi(x,y)$ for $H=4$\\
                 with $a=b=0.5$, $d=1.0$, $N=0.2$, $m=0.1$, $\phi=0.0$, $\theta=2.4$, $R_{m}=1.0$\\

   \includegraphics[width=3.8in,height=2.4in ]{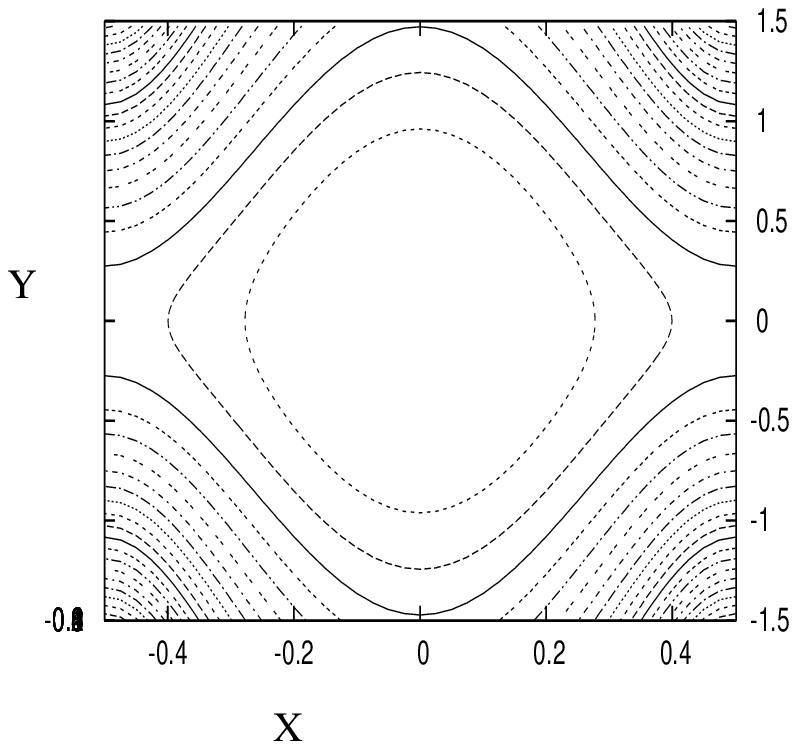}\\
 Fig. 31   Distribution of magnetic force function $\phi(x,y)$ for $H=6$\\
                 with $a=b=0.5$, $d=1.0$, $N=0.2$, $m=0.1$, $\phi=0.0$, $\theta=2.4$, $R_{m}=1.0$\\
\end{center}
\end{minipage}\vspace*{.5cm}\\

\begin{minipage}{1.0\textwidth}
   \begin{center}
\includegraphics[width=3.8in,height=2.4in ]{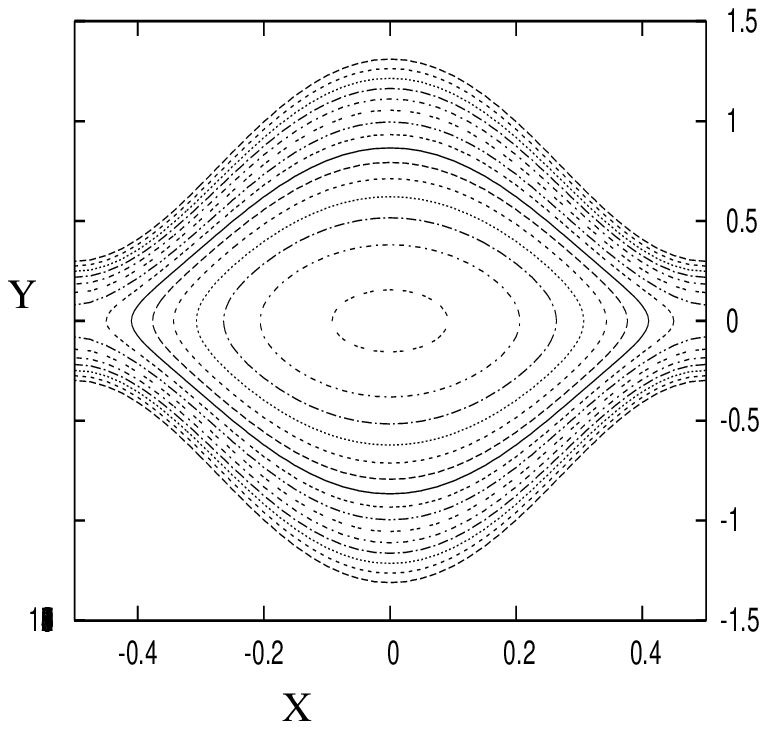}\\
Fig. 32 Distribution of magnetic force function $\phi(x,y)$ for $N=0.4$\\
                with $a=b=0.5$, $d=1.0$, $H=2$, $m=0.1$, $\phi=0.0$, $\theta=2.4$, $R_{m}=1.0$\\

\includegraphics[width=3.8in,height=2.4in ]{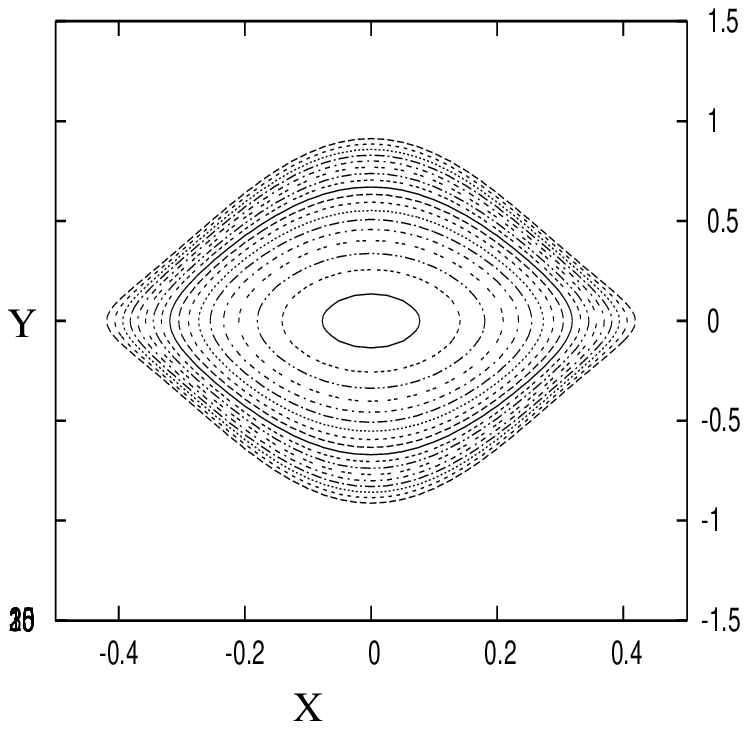}\\
Fig. 33 Distribution of magnetic force function $\phi(x,y)$ for $N=0.8$\\
                 with $a=b=0.5$, $d=1.0$, $H=2$, $m=0.1$, $\phi=0.0$, $\theta=2.4$, $R_{m}=1.0$\\

\includegraphics[width=3.8in,height=2.4in ]{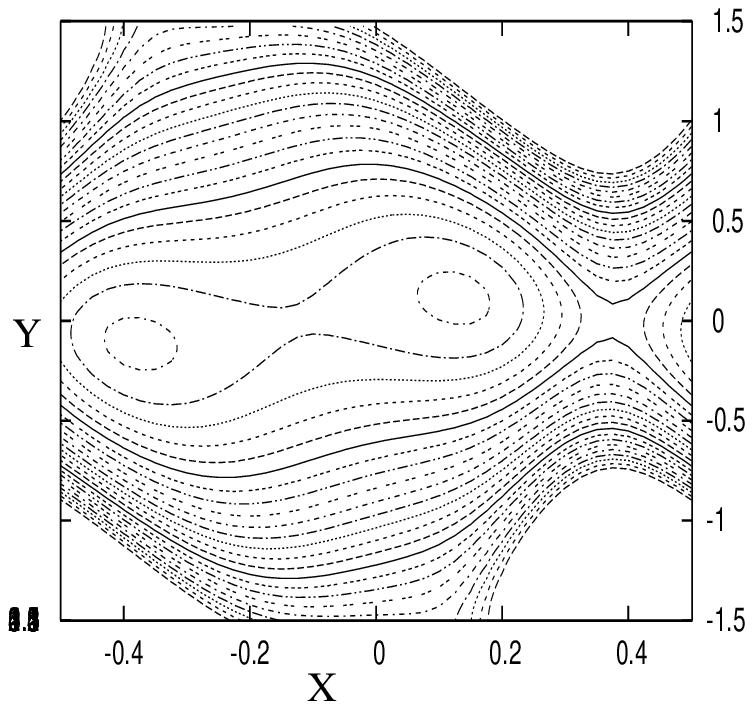}\\
Fig. 34 Distribution of magnetic force function $\phi(x,y)$ for $\phi=\pi/2$\\
                 with $a=b=0.5$, $d=1.0$, $H=2$, $m=0.1$, $N=0.2$, $\theta=2.4$, $R_{m}=1.0$\\
\end{center}
\end{minipage}\vspace*{.5cm}\\

\begin{minipage}{1.0\textwidth}
   \begin{center}
   \includegraphics[width=3.8in,height=2.4in ]{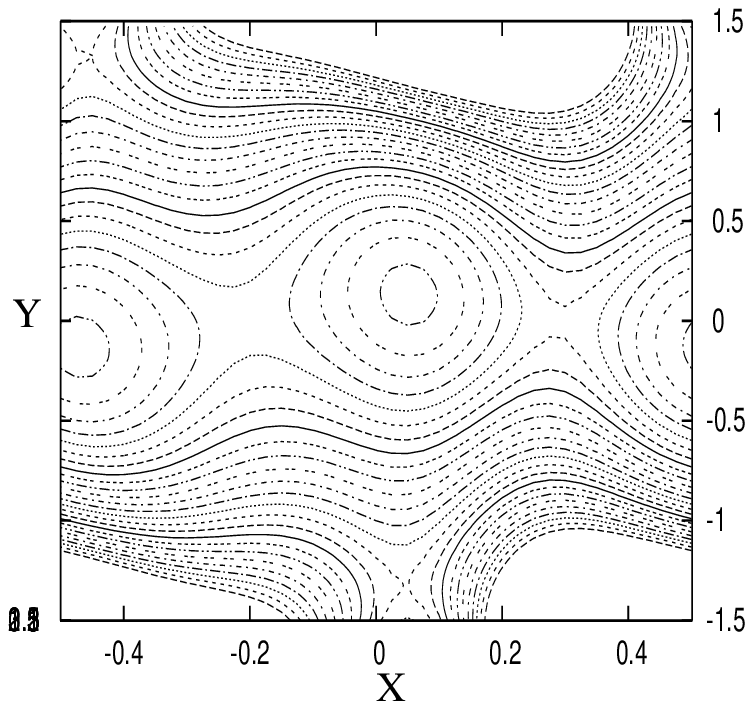}\\
 Fig. 35  Distribution of magnetic force function $\phi(x,y)$ for $\phi=5\pi/6$\\
                 with $a=b=0.5$, $d=1.0$, $H=2$, $m=0.1$, $N=0.2$, $\theta=2.4$, $R_{m}=1.0$\\

\includegraphics[width=3.8in,height=2.4in ]{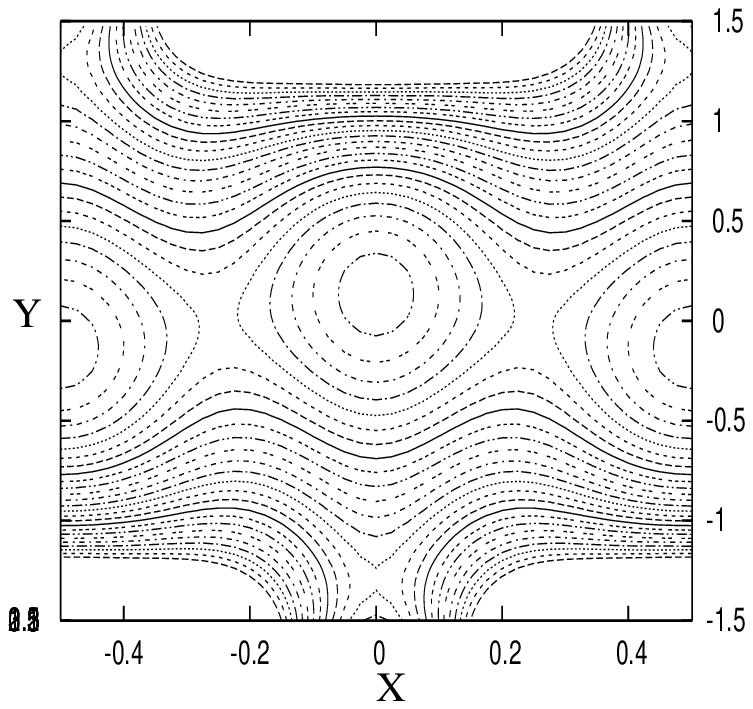}\\
Fig. 36 Distribution of magnetic force function $\phi(x,y)$ for $\phi=\pi$\\
                 with $a=b=0.5$, $d=1.0$, $H=2$, $m=0.1$, $N=0.2$, $\theta=2.4$, $R_{m}=1.0$\\
\end{center}
\end{minipage}\vspace*{.5cm}\\


\begin{thebibliography}{99} \singlespacing
\bibitem {R1} Latham, T. W.: Fluid motion in a peristaltic pumps,
 M. S. Thesis,MIT, Cambridge, M. A., 1966.
\bibitem {R2} Fung, Y. C. and Yih, C. S. : Peristaltic transport,
 Trans ASME J. Appl. Mech. 35 (1968) 669-675.
\bibitem {R3} Burns, Y. C. and Perkes, T. : Peristaltic Motion,
 J. Fluid Mech. 29 (1967) 731-743.
\bibitem {R4} Chow, T. S. : Peristaltic transport in a circular
 cylindrical pipe, Trans ASME J. Appl. Mech. 37 (1970) 901-905.
\bibitem {R5} Hung, T. K. and Brown T. D. : Solid-Particle Motion in two
 dimensional peristaltic flows, J. Fluid Mech., 73(1976) 77-96
\bibitem {R6} Takabatake, S. and Ayukava, K. : Numerical study of
 two dimensional peristaltic flows, J. Fluid Mech. 122 (1982) 439-465.
\bibitem {R7} Srivastava, L. M. and Srivastava, V. P. : Peristaltic
 transport of a particle-fluid suspension, Trans. ASME J. Biomech.
 Engg. 111 (1989) 157-165.
\bibitem {R8} Misra, J. C. and Pandey, S. K. : Peristaltic
 transport of a particle-fluid suspension in a cylindrical tube,
  Comput. Math. Applic. 28 (1994) 131-145.
\bibitem {R9} Antanovskii, L. K. and Ramkisson, H. : Long-Wave
 Peristaltic transport of a compressible viscous fluid in a finite
  pipe subject to a time dependent pressure drop, J. Fluid Dyn.
  Research, 19 (1997) 115-123.
\bibitem {R10}  Misra, J. C. and Pandey, S. K. : Peristaltic
 transport of blood in small vessels: Study of mathematical model,
 Comput. Math. Applic. 43 (2002) 1183-1193.
\bibitem {R11} Srivastava, L. M. and Srivastava, V. P. : Effects of
 poiseuille flow on peristaltic transport of a particulate suspension,
 Z. Angew. Math. Phys. (ZAMP), 46 (1995) 655-679.
\bibitem {R12} Mekheimer, Kh. S.; El Shehawey, E. E. and Elaw, A. M.
: Peristaltic motion of a particle-fluid suspension in a planar
channel., Int. J. Theo. Phys. 37 (1998)2895-2920.
\bibitem {R13} Srivastava, L. M. and Srivastava, V. P. : Peristaltic
  transport of Power-law fluid : Application to the ductus of
  efferentes of the reproductive tract, Rheol. Acta, 27 (1988)  428-433
\bibitem {R14} Usha, S. and Rao, R. : Perisataltic transport of
 two-layered Power-law fluids, Trans ASME J. Biomech.
  Engg. 199 (1997) 483-488.
\bibitem {R15} Lee, J. S. and Fung, Y. C. : Flow in a non-uniform
 small blood vessels, Microvasc. Res. 3 (1971) 272-287.
\bibitem {R16} Carew, F. O. and Pedley T. J. : An active membrane
 model for peristaltic pumping: Part-1-periodic activation waves
 in an infinite tube, J. Biomech. Eng. 119 (1997) 66-76
\bibitem {R17} Taylor, G. I. : Analysis of the swimming of
 microscopic organisms, Proc. R. Soc. Lond. A 209 (1951) 447-467.
\bibitem{R18} Mishra, M. and Rao, A. R. : Peristaltic transport
 of a Newtonian fluid in an asymmetric channel, Z. Angen.
 Math. Phys. (ZAMP) 54 (2003) 532-550
\bibitem {R19} Rao, A. R. and Mishra, M.: Non-linear and
 curvature effects on peristaltic flow of a viscous fluid in
  an asymmetric channel, Acta Mech. 168 (2004) 35-59
\bibitem {R20} Reddy, M. V. S., Rao, A. R. and Sreenadh, S. :
  Peristaltic transport of a Power-law fluid in an asymmetric channel,
 Int. J. Non-linear Mech., 42 (2007) 1153-1161.
\bibitem {R21} Eytan, O. and Elad, D. : Analysis of intra-Urterine
 fluid motion induced by uterine contractions, Bull. Math.
  Biology 61 (1999) 221-238.
\bibitem {R22} Erigen, A. C. : Theory of micropolar fluids,
 J. Math. Mech. 16 (1966) 1-18.
\bibitem {R23}  Ariman, T., Turk, M. A. and Sylvaster, N. D. : Review
 article - applications of microcontinuum fluid mechanics,
  Int. J. Eng. Sci., 12 (1974) 273-293.
\bibitem {R24} Ariman, T., Turk, M. A. and Sylvaster, N. D. : On
 the steady and peristaltic flow of blood,
  J. Appl. Mech. 41 (1974) 1-7.
\bibitem {R25} Srinivasacharya, D. Mishra, M. and Rao, A. R. :
 Peristaltic pumping of a micropolar fluid in a tube,
  Acta. Mech., 161 (2003) 165-178.
\bibitem {R26} Lukaszewicz, G.: Micropolar fluids -
 Theory and applications, Birkhauser, 1999.
\bibitem {R27} Mathu, P., Rathiskumar, B. V. and Chandra, P. : On
 the inference of wall properties in the peristaltic motion of
  micropolar fluid, ANZIAM J. 45 (2003) 245-260.
\bibitem{R28} Mathu, P., Rathiskumar, B. V. and Chandra, P. :
 Peristaltic motion of micropolar fluid in circular
  cylindrical tubes: Effect of wall properties,
   Appl. Math. Model. 32 (2008) 2019-2033.
\bibitem {R29} Ali, N and Hayat, T. : Peristaltic flow of
 a micropolar fluid in an asymmetric channel,
  Copmut. Math. Applic. 55 (2008) 589-608.
\bibitem {R30} Hayat, T. and Ali, N. : Effects of an endoscope on
peristaltic flow of a micropolar fluid, Math. Comput. Model. 48
(2008) 721-733.
\bibitem {R31}  Agarwal, H. L. and Anwaruddin, B. : Peristaltic
 flow of blood in a branch, Ranchi Univ. Math. J., 15 (1984) 111-121.
\bibitem{R32} Li, A.; Nesterov, N. I., Malikova, S. N. and
 Kiiatkin, V. A. : The use of an impulsive magnetic field in the
 combined therapy of patients with stone fragnents in the upper
 urinary tract, Vopr. Kurortol Fizioter Lech Fiz Kult, 3 (1994) 22-24.
\bibitem {R33} Mekheimer, Kh.S. and Al-Arabi, T. H. : Non-linear
 peristaltic transport of MHD flow through a porous medium,
  Int. J. Math. Sci. 26 (2003 1663-1682.
\bibitem{R34} Misra, J. C., Maiti, S. and Shit, G. C. : Peristaltic
 transport of a physiological fluid in an asymmetric porous channel
  in the presence of an external magfnetic field,
   J. Mech. Med. Biol. 8 (2008) 507-525.
\bibitem{R35}  Elashahed, M and Harun, M.H. : Peristaltic transport
 of Johnson-Segalman fluid under effect of a magnetic field,
  Math. Problems Eng. 2005 (6) (2005) 663-677.
\bibitem {R36}  Vishnyakov, V. I. and Pavlov K. B. : Peristaltic
 flow of a conductive liquid in a transverse magnetic field,
   Magnetohydrodyanamics, 8 (1972) 174-178.
\bibitem {R37}  Mekheimer, Kh. S. : Effect of the induced
 magnetic field on peristaltic flow of a
 couple stress fluid, Phys. Letters A. 372 (2008) 4271-4278.
\bibitem {R38} Mekheimer, Kh. S. : Peristaltic flow of a
 magneto-micropolar fluid: Effect of induced magnetic field;
 J. Appl. Math. vol.-2008 (2008) 570825, 23 Pages.
\bibitem {R39} Shapiro, A.H.; Jafrin, M.Y. and Weinberg, S.L.:
Peristaltic pumping with long wavelengths at low Reynolds number,
J. Fluid Mech. 37(1969) 799-825
\end{thebibliography}
\end{document}